\documentclass[journal]{IEEEtran}
\usepackage{stmaryrd}
\usepackage{cite}
\usepackage{float}
\usepackage{amsmath,amssymb,amsfonts}
\usepackage{algorithmic}
\usepackage{graphicx} 
\usepackage[table]{xcolor}
\usepackage{textcomp}
\usepackage{xcolor}
\usepackage{booktabs}
\def\BibTeX{{\rm B\kern-.05em{\sc i\kern-.025em b}\kern-.08em
    T\kern-.1667em\lower.7ex\hbox{E}\kern-.125emX}}
    \usepackage{caption}
\usepackage{subfigure}
\usepackage{mathtools}
\usepackage{multirow}
\usepackage{lettrine}
\usepackage[ruled,vlined,linesnumbered]{algorithm2e}
\usepackage{babel,blindtext}
\usepackage[spaces,hyphens]{url}

\usepackage{tabularx}
\usepackage[justification=centering]{caption}
\usepackage{amsthm}

\usepackage{nccmath}
\usepackage{booktabs,tikz}

\usepackage{lipsum, babel}
\ifCLASSINFOpdf
 \else
\fi

\begin{document}

\title{Age of Information Optimization in  RIS-Assisted Wireless Networks}

 \author{ Ali~Muhammad, Mohamed Elhattab, Mohamed Amine Arfaoui, Ahmed Al-Hilo,
         and Chadi Assi \thanks{A part of this paper will be presented in IEEE International Conference on Communications (ICC), South Korea, May 2022 \cite{aliicc2022}, where the objective is to  minimize the sum AoI and the problem is solved on a time-slot basis. In this article, and different from \cite{aliicc2022}, the AoI minimization problem is solved with the  objective of minimizing the expected sum AoI  over a time horizon.}
\thanks{Ali Muhammad and Mohamed Elhattab are with the Department of Electrical and Computer Engineering, Concordia University, Montreal, QC H3G 1M8, Canada}
\thanks{Mohamed Amine Arfaoui, Ahmed Al-Hilo and Chadi Assi are with the Concordia Institute of Information System
Engineering, Concordia University, Montreal, QC H3G 1M8, Canada}
}

%\markboth{Journal of \LaTeX\ Class Files,~Vol.~14, No.~8, January~2021}%

\maketitle
\begin{abstract}
In this paper, we consider a wireless network consisting of a base station (BS) that is serving multiple real-time traffic streams forwarding information updates to their destinations in order to sustain the freshness of information for time-critical applications. Since the wireless channels may be unreliable due to the impurities of the propagation environments, such as deep fading, blockages, etc., we integrate a reconfigurable intelligent surface (RIS) to the wireless system in order to mitigate the propagation-induced impairments, enhance the quality of the wireless links, and ensure that the required freshness of information is achieved for these real time applications. For this network set-up, we investigate the joint optimization of the traffic streams scheduling and the RIS phase-shift matrix with the goal of minimizing the long-term average Age of Information (AoI). The formulated optimization problem is a mixed integer non-convex optimization problem, which is difficult to solve. To circumvent the high-coupled optimization variables, and with the aid of bilevel optimization, we decompose the original problem into an outer traffic stream scheduling problem and an inner RIS phase-shift matrix problem. For the outer problem, owing to its complexity and stochastic nature of packet arrivals, we resort to deep reinforcement learning (DRL) solution where the traffic stream scheduling is modeled as a Markov Decision Process (MDP), and Proximal Policy Optimization (PPO) is invoked to solve it. Whereas, the inner problem that determines the RIS configuration is solved through semi-definite relaxation (SDR). Finally, we show through extensive simulations that our approach evaluates the combined impact of scheduling policy and RIS configuration on the long term average AoI, where we demonstrate its superiority against other baseline schemes.

\end{abstract}

\IEEEpeerreviewmaketitle

\section{Introduction}
\subsection{Motivation}
\IEEEPARstart{N}{ext} generation wireless networks (5G and Beyond, 6G) aim to provide tremendous improvements over previous generations by offering a massive connectivity, ultra reliable and low-latency communications, and soaring broadband speeds. Such transformation will give rise to a wide range of propitious applications such as intelligent transportation systems (ITS), tactile internet, augmented/virtual reality, industry 4.0, etc. The crux of these applications is critical decisions that rely on real-time information updates. For example, a Cooperative Autonomous Driving (CAD) system is an ITS application wherein status information, such as speed and vehicle position, along with other sensory data, are crucial to be timely disseminated for safety reasons. Another application pertains to traffic monitoring and control systems, where several security cameras monitor the traffic, and in case of any accident, it must immediately inform the control center for a quick dispatch of emergency vehicles. In the above examples, if the information delivered is not "fresh", i.e., if the information updates are not timely delivered, there may be severe consequences impacting not only the performance of these intelligent and critical systems, but also the safety and wellness of people. Thus, reliability and timeliness in delivering status updates are of primordial importance for these real-time applications. \\ 
\indent Recently, information freshness has been investigated through defining a new performance metric that is Age of Information (AoI). AoI quantifies the freshness of status updates from the destination perspective \cite{9215013}. AoI is defined as the elapsed time since the most recent delivered status message was generated \cite{kosta2017age}. AoI has brought a sheer novelty in specifying the information freshness against other metrics such as delay and latency for time-critical applications, and hence, it has been widely studied recently \cite{yates2021age}. In reality, the timely delivery of information update messages is challenging due to the behaviour of wireless communication environments, which may be highly random and uncontrollable. Typically, a strong communication link between a source and destination cannot be guaranteed due to channel impairments and blockages. Thus, the question that arises here is the following: \textit{how can one guarantee reliable wireless communication links in highly random and uncontrollable environments?} Reconfigurable intelligent surfaces (RIS) technology has been envisioned as a key solution which provides the answer to the above question.\\
\indent Principally, a potential solution to circumvent the impairments of the wireless propagation environments and construct a strong channel between the source and destination is to recognize alternative propagation routes through which the information-bearing signal can be received at the point-of-interest. This can be achieved by using the RIS technology. RIS has been recently proposed as a new paradigm that will enable the next-generation wireless networks \cite{9122596}. It has received great attention from the academic and industrial research communities due to its potential capability in improving the wireless links' quality by reshaping and re-configuring the wireless environment \cite{subrt2012intelligent}. Specifically, RIS consists of an array of passive elements, where each element has the ability to independently tune the phase-shift of the impinging waves. Therefore, the signals transmitted within the wireless propagation environments can be controlled, and through a proper adjustment of the phase shifts of all the RIS elements, the desired signals at the points of interest can be enhanced \cite{Marco_JSAC}. It is worth mentioning that the benefits of introducing RIS to enhance the QoS performance of communication networks have been unveiled in the literature for real-time applications. For example, in vehicle-to-everything (V2X) applications \cite{chen2021qos}, RIS has proven to improve the QoS in harsh transmission environments. Similarly, in smart industry application \cite{dhok2021non}, RIS offers promising signal strength and quality over longer distances. Motivated by these facts, it is foreseen that the information freshness can be significantly improved through the integration of RIS, especially for time-sensitive applications and services, which is indeed the focus of this paper.

% in has been for a broad range of practical applications. For example, However, 
% in this study is aimed to improve the channel
% quality between the Base Station (BS) and a set of destinations in order to enhance the freshness
% of information. Conversely, a poor channel quality may directly impact the transmission of
% status updates which may result in erroneous decisions at the destinations.

%induces a certain phase-shift independently to the incident electromagnetic waves through a controller
%
\subsection{Related Works}
 \label{relatedwork}
The aim of this work is to investigate the AoI improvement that can be brought by RIS. Based on this, the two main components of this study are AoI and RIS. Here, we present the relevant works related to AoI and RIS that are reported in the literature.
\subsubsection{AoI based data transmissions}
AoI metric has received a considerable interest from the research community, accentuating its benefits especially for time-sensitive systems. Different from traditional performance metrics, such as delay and throughput, AoI captures the freshness of information through the inter-delivery time intervals of the packets as well as the delay experienced by the packets in the system.  
The AoI minimization problem has been investigated in various domains, such as, vehicular networks \cite{alabbasi2020joint,vehicular1}, machine-type communication \cite{9126228}, UAV-assisted communications \cite{8570843,9511090},  edge caching \cite{9262040,edgecaching2}, and mobile edge computing assisted networks \cite{9001219,9665756}. More relevant to this research are the works that have investigated the AoI minimization problem with stochastic arrivals. In \cite{kadota2019minimizing}, a lower bound on the average AoI performance is derived for networks with stochastic packets arrivals under three different queuing scenarios, namely, no queue, single queue, and first-in first-out queue. The authors of \cite{9007478} investigated the AoI in a carrier-sense multiple access based system employing the stochastic hybrid system tools where $N$ links contend for a channel. %\textcolor{blue}{stochastic hybrid systems (define briefly this technique)}. 
They aimed to optimize the average AoI by adjusting the back-off time of each link. In \cite{8437712}, the author proposed a near-optimal solution to address the optimization of AoI in wireless communication networks wherein Whittle’s index was used to capture the transmission urgency of terminals. The authors of \cite{li2019minimizing} considered various sampling periods and sample sizes for each source node and proposed a low-complexity scheduling algorithm that achieves near-optimal performance when there is no synchronization among the nodes during the sampling process. The authors of \cite{zhang2020age} investigated the AoI minimization problem in the context of cellular Internet of UAVs and formulated a framework that jointly optimize the sensing and transmission time, the UAV trajectory and the task scheduling (i.e., the selection of the sensing tasks). %\textcolor{blue}{scheduling (packets or users?)}. 
The formulated problem, which is NP-hard, was decoupled into two sub-problems and that were solved using an iterative algorithm and a dynamic programming approach, respectively.
\par Different from the above background, this work leverages RIS in a wireless network to enhance information freshness at the end users by minimizing the AoI. In the following part, we will discuss the recent research contributions on the integration of RIS in wireless cellular networks.
\begin{table}[t]
          \centering
      \caption{Table of Notations.}
     \begin{tabular}{|c|l|}
     \hline
           \textbf{Parameters}  & \textbf{Description}  \\ \hline
        $T$  & Time horizon of the discrete time system.  \\ \hline
          $\mathcal{T}$  & Set of time slot indices within the time interval $[1,T]$.  \\ \hline
         $I$ & Number of traffic streams. \\ \hline
          $\mathcal{I}$ & Set of traffic streams. \\ \hline
      $N$ & Number of available channels resources. \\ \hline
      $\mathcal{N}$ & Set of the $N$ channels. \\ \hline
          $\lambda_i$ & Probability with that a packet from  \\
          & stream $i$ arrives to the system. \\ \hline
          $u_{i}(t)$ & Indicator that a packet from traffic stream \\
          &  $i$ arrives in slot $t$ in the queue. \\ \hline
            $x_{i,n}(t)$ & Binary variable equals to 1 when a packet \\
          & from traffic stream $i$ is scheduled in slot $t$ on channel $n$. \\ \hline
            $z_{i}(t)$ & System time of the  packet in $Q_i$ of user  \\
          & stream $i$ at the beginning of slot $t$.\\   \hline
            $\beta_{i}(t)$ & Indicator with value 1 if the selected \\
          & stream has a non-empty queue.\\    \hline 
            $y_{i}(t)$ & The age corresponding to destination $i$ in time-slot $t$.\\    \hline
            $\Phi_F(t)$ &  Phase-shift for $f$th reflecting element \\ \hline
        $\hat {\boldsymbol h}_{b \rightarrow R,n}(t)$ & Small-scale fading between the BS and RIS \\ \hline  $\hat {\boldsymbol h}_{R \rightarrow i,n}(t)$ &  Small-scale fading between the RIS and destination $i$ \\ \hline 
        % $\Delta_{b \rightarrow R,n}(t)$ & path-loss coefficient between the  BS and RIS \\ \hline 
        % $\Delta_{R \rightarrow i,n}(t)$ &path-loss coefficient between the RIS and destination $i$ \\ \hline 
        $\gamma_{\rm th}$ & Threshold to ensure reliable decoding \\ \hline      \end{tabular}
         \label{notations}
 \end{table}
 \subsubsection{RIS aided Wireless Networks}
In \cite{8811733}, the authors addressed the minimization problem of total transmit power at the transmitter by jointly optimizing the transmit beamforming through the active antenna array of the transmitter and the passive beamforming through the phase-shift elements of the RIS. The authors of \cite{9206044} developed different free-space path-loss models for RIS-assisted wireless communications, with the goal of enhancing the network coverage in a cost-effective and energy-efficient way through optimizing the phase-shifts of the RIS elements. Considering the potential challenges pertaining to spectrum and energy usage in Device-to-Device (D2D) communication, the authors in \cite{9301375} focused on an RIS-assisted uplink D2D-enabled cellular networks and investigated the joint power allocation  and RIS phase-shift optimization problem with an objective to maximize the sum rate. 
%\textcolor{blue}{what was the metric? sum rate? minimum rate?}.
The authors of \cite{9211520} investigated the resource allocation problem for multi users communication leveraging the RIS. More specifically, the total transmit power is minimized through an optimal design of the transmit power at the base station (BS) and the passive beamforming at the RIS. In \cite{9138463}, the authors proposed a two-way communication model assisted by an RIS, where the objective was to maximize the minimum received signal-to-interference-plus-noise ratio (SINR) at the cellular users by optimizing the RIS configuration. The paramount security performance of multi-input and multi-output wireless communication systems is probed by invoking the RIS in \cite{9201173}, where the aim was maximizing the secrecy rates through a proper design of the RIS configuration and the transmit power. However, none of these works studied the effect of the RIS on improving the AoI. 
\par Recently, the authors of \cite{samir2020optimizing} addressed the AoI minimization problem in UAV-assisted RIS networks, where the objective was to minimize the average sum AoI by optimizing the altitude of UAVs, the RIS configuration and the scheduling decisions. Although the work in \cite{samir2020optimizing} is the first that studied the AoI minimization problem through the use of RIS, it considered the scheduling of only a single user within a given time-slot and ignored the direct channels from the BS to the users. To the best of our knowledge, the integration of RIS in time-sensitive applications, where the freshness of information is of critical importance, is still far from being mature. Motivated by this, we consider, in this work, the problem of scheduling a finite number of streams in order to transmit their information update messages, where, as opposed to  \cite{samir2020optimizing}, we considered a more general setting by including the direct channels from the BS to the users. 
\subsection{Contributions and Outcomes}
{ In this paper, we investigate the AoI minimization problem in RIS-aided time-sensitive applications. Specifically, we aim at minimizing the expected sum AoI by optimizing the user scheduling decisions and the phase shifts of RIS elements. The main contributions of this work are summarized as follows:
\begin{itemize}
    \item We formulate a joint user scheduling and phase-shift matrix (passive beamforming) optimization problem with the objective of minimizing the expected sum AoI of multiple traffic streams.
    \item Owing to the stochastic nature of arrival of packets, the combinatorial nature of the user scheduling task, and the non-convexity of the different system constraints, it is extremely challenging to solve the formulated problem. Alternatively, with the aid of bi-level optimization, the original problem is reformulated into an outer user scheduling problem and an inner phase-shift matrix optimization problem. Owing to the complexity and stochastic nature of the packet arrivals, the outer problem is formulated as a Markov Decision Process (MDP) and solved through the Deep Reinforcement Learning (DRL) technique. For the inner problem, an efficient algorithm based on semi-definite relaxation (SDR) is proposed.
    \item The performance of the proposed approach is assessed through extensive simulations, where different baseline methods were considered for comparison purposes. We demonstrate that our proposed scheme achieves the minimum expected sum AoI in contrast with the other considered methods.
\end{itemize}
\indent In the simulation results, we show how the integration of RIS can significantly reduce the AoI of time-critical applications as compared to the case where there is no RIS. In addition, for comparison purposes, three baseline schemes, namely, greedy scheduling with SDR, Round-Robin scheduling with SDR and DRL with a random RIS configuration, were adopted and we demonstrated the superiority of the proposed scheme. Finally, we evaluate the performance of the proposed scheme and baseline approaches with respect to different system parameters, including the size and the location of RIS and the arrival rate of the updates' packets.}
\subsection{Outline and Notations}
{The remainder of this paper is organized as follows. The system model is presented in Section~\ref{sysmodel}. The definition of age of information accompanied by an illustrative example is presented in Section~\ref{aoiDefinition}. The problem formulation, the proposed scheduling and phase shift optimization algorithms are explained in Section~\ref{formulationprob}.
Section~\ref{simulationresults} demonstrates the performance evaluations on the proposed algorithm with different parameter settings. Finally, Section \ref{conclusion} concludes the paper. The notations used throughout the paper are summarized in Table \ref{notations}.}

\section{System Model}
\label{sysmodel}
\subsection{Network Model}
\begin{figure}[t]
\centering
\includegraphics[width=1\linewidth]{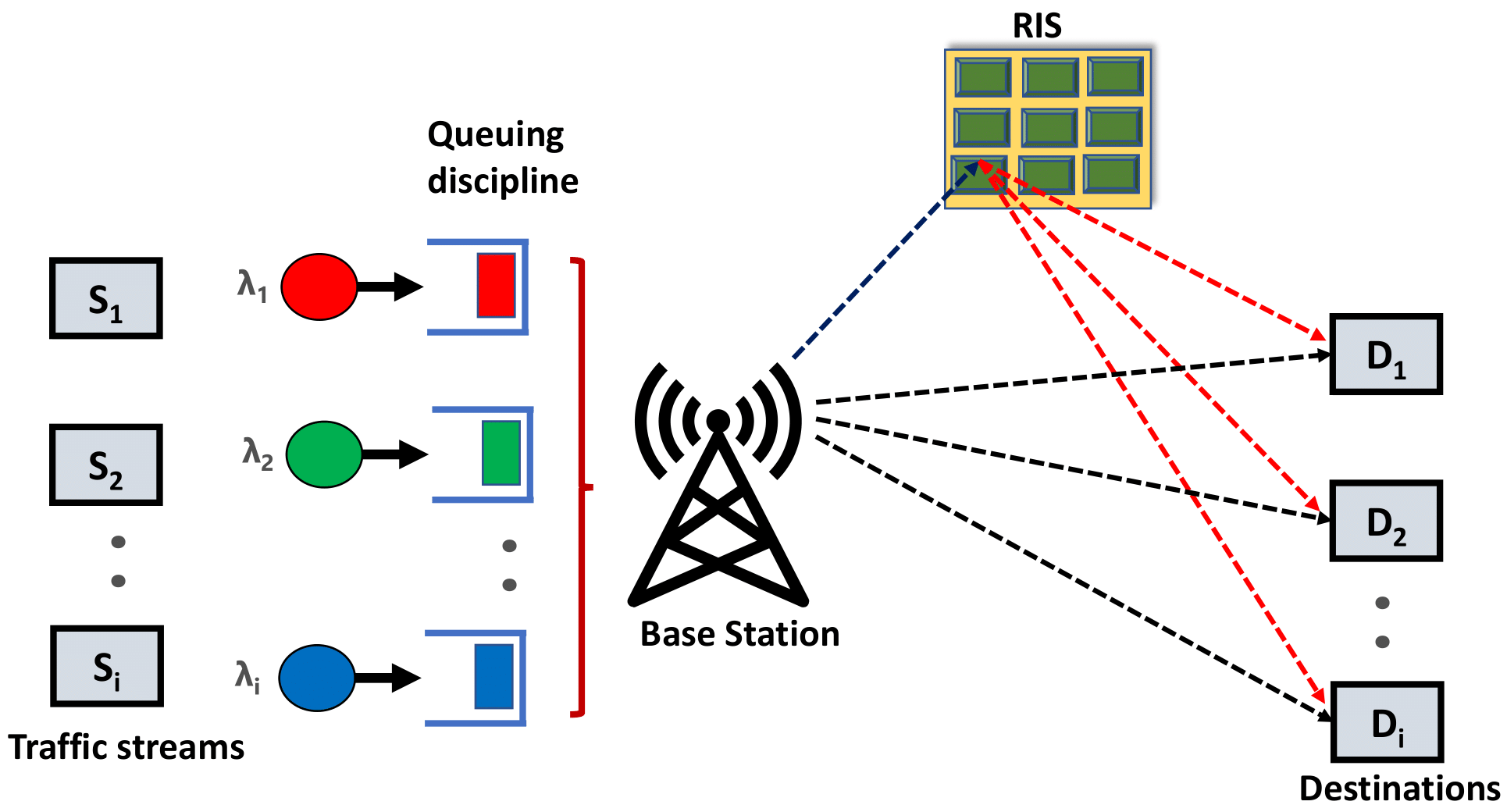}
\caption{An illustration of our system model.}
\label{systemmodel}
\end{figure}

We consider a downlink wireless network consisting of one base station (BS), equipped with a single antenna, that is serving $I$ traffic streams to forward their status-update messages to $I$ destinations as depicted in Fig. \ref{systemmodel}.\footnote{{In this work, we focus on studying the fundamentals and presenting a proof of concept for RIS-enabled single-input-single-output (SISO) wireless networks that generate real-time information updates, where our main target is characterizing the corresponding performance, in terms of AoI, in order to provide succinct insights. The use of multiple antennas at the BS can significantly boost the performance of the proposed model. This will be considered in future works, where the proposed techniques in the current work can be exploited.}} We assume that the BS is equipped with $I$ virtual queues, within which the BS only stores the most recent packet of each stream. The time dimension is slotted into time-slots, where each is represented by a time-slot index $t \in [1,\infty)$. Let $T$ denote the time horizon of this discrete-time system. In addition, let $\mathcal{T} = \{1, 2, \dots , T\}$ denote the set of time slot indices within the time interval $[1,T]$ and let  $\mathcal{I} = \{1, 2, \dots ,  I\}$ denote the set of the traffic streams. In this setting, at the beginning of every time-slot $t\in \mathcal{T}$, a packet from stream $i \in \mathcal{I}$ arrives to the system with a probability $\lambda_i~\in(0,1]$. Accordingly, for all $t\in \mathcal{T}$ and $i \in \mathcal{I}$, let $u_{i}(t)$ be the binary variable that indicates whether a packet from the $i$th traffic stream arrives to the BS at the $t$th time-slot or not. Based on its definition, for all $i \in \mathcal{I}$, the arrival process $u_i(t)$ is a Bernoulli arrival process that is i.i.d over time, where $P(u_{i}(t)=1) =  \lambda_i$. Moreover, the arrival processes $\left(u_i(t) \right)_{1 \leq i \leq I}$ are independent across the different streams. 

\par Due to impurities and the obstacles of the wireless propagation environment, the existence of a strong direct line-of-sight (LoS) communication link between the BS and each destination is difficult to obtain. For this purpose, an RIS equipped with $F$ reflecting elements is assumed to be deployed within the considered wireless network to assist the transmission from the BS by passively relaying the status update information to the destinations. The BS continuously controls the phase-shift of the reflecting elements in order to maintain the quality of service (QoS) required by the destinations. In this context, for all $t\in \{1, 2, \dots , T\}$, let $ \boldsymbol \Phi(t)=  {\rm diag} \left( \exp \left[\boldsymbol{\theta}(t)\right] \right) \in \mathbb{C}^{F \times F}$ denotes the $F \times F$ phase-shift matrix of the RIS, where $\boldsymbol \theta(t) =[  \theta_1(t),\theta_2(t),\dots,\theta_F(t)]^T$ is the $F\times 1$ vector that contains the phase-shifts of the RIS, such that, for all $f \in \mathcal{F} \triangleq \left\{ 1,2,\dots,F\right\}$,  $\theta_f(t) \in [0,2\pi)$ is the phase-shift of the $f$th reflecting element of the RIS. 
\par {The total bandwidth available at the BS is divided into $N$ channels, where each channel consists of one spectrum resource. The channel diversity exists between different channels and BS can schedule different traffic streams to at most $N$ channels where each traffic stream is assumed to be allocated to only one channel~\cite{li2019minimizing},\cite{8340064}. Let $\mathcal{N} = \{ 1, 2, ....N \}$ denote the set of the $N$ channels.} {Moreover, for all $t\in \mathcal{T}$, $i \in \mathcal{I}$ and $n \in \mathcal{N}$, let $x_{i,n}(t)$ be the indicator whether the $i$th stream has been scheduled by the BS on $n$th channel in the $t$th time-slot or not \footnote{Note that the transmission from BS to each of the destinations through each direct link (BS-destination) and indirect link (BS-RIS-destination) takes only one time-slot. It is worth mentioning that the RIS is a full-duplex technology with interference-free transmission.}. This is to note that the term scheduling is collectively used for selecting a traffic stream and allocating it a channel for transmission. On the other hand, scheduling a traffic stream without allocating a channel resource and vice versa have no meaning at all. $x_{i,n}(t)$ is defined as follows: } 

\begin{equation}
  x_{i,n}(t) =
 \begin{cases}
  {1} & \text{if traffic stream $i$ is scheduled}  \\
 &  \text{ on channel $n$ in time-slot $t$},
  \\
  0 &\text{otherwise,}
 \end{cases}
 \label{mainage}
\end{equation}

{Based on this, the transmission scheduling constraints are given as follows.}

{
\begin{align}
&\sum_{i=1}^ {I} \sum_{n=1}^ {N} x_{i,n}(t)  \leq N, \qquad \,\,\,\,\, \forall t \in \mathcal{T}. \label{cons_channel1}  \\
&\sum_{n=1}^ {N} x_{i,n}(t)  \leq 1, \qquad \forall t \in \mathcal{T}, i \in \mathcal{I}. \label{cons_channel2}
\end{align}}
\subsection{Channel Model and SNR Analysis}
\par {For all $t \in \mathcal{T}$, $i \in \mathcal{I}$, and $n \in \mathcal{N}$, the channel coefficients between the BS and the RIS, between the RIS and the $i$th destination, and between the BS and the $i$th destination on the $n$th spectrum resource are denoted, respectively, by $\boldsymbol h_{b \rightarrow R,n}\boldsymbol(t) \in \mathbb{C}^{F \times 1}$, $\boldsymbol h_{ R \rightarrow  i,n}(t) \in \mathbb{C}^{F \times 1}$ and $h_{b \rightarrow i,n}(t)  \in \mathbb{C}$. All channel coefficients consist of both the small-scale fading and the large-scale fading. In fact, for all $t \in \mathcal{T}$, $i \in \mathcal{I}$, and $n \in \mathcal{N}$, the channel coefficients $\boldsymbol h_{ b \rightarrow  R,n}(t)$, $\boldsymbol h_{ R \rightarrow  i,n}(t)$ and $h_{b \rightarrow i,n}(t)$ are expressed, respectively, as}
{
\begin{align}
&\boldsymbol h_{ b \rightarrow  R,n}(t) =   \hat {\boldsymbol h}_{ b \rightarrow  R,n}(t) \Delta_{b \rightarrow R} \label{gain1}  \\
& \boldsymbol  h_{ R \rightarrow  i,n}(t) =  \hat {\boldsymbol h}_{ R \rightarrow  i,n}(t) \Delta_{R \rightarrow i} \label{gain2} \\ 
&h_{b \rightarrow i,n}(t) = \hat h_{b \rightarrow i,n}(t) \Delta_{b \rightarrow i} \label{gain21}
\end{align}}
\noindent {where $\hat { \boldsymbol h}_{ b \rightarrow  R,n}(t)$, $\hat { \boldsymbol h}_{ R \rightarrow  i,n}(t)$ and $\hat {\boldsymbol h}_{ b \rightarrow  i,n}(t)$ represent the small-scale fading coefficients between the BS and the RIS, between the RIS and the $i$th destination, and between the BS and $i$th destination on the $n$th frequency resource, respectively, whereas $\Delta_{b \rightarrow R}$, $\Delta_{R \rightarrow i}$ and  $\Delta_{b \rightarrow i}$ represent the large-scale fading coefficients between the BS and RIS, between the RIS and the $i$th destination, and between the BS and the $i$th destination respectively. Additionally, for all $i \in \mathcal{I}$, and $n \in \mathcal{N}$, the large-scale fading coefficients can be modeled as}
{
\begin{align}
&\Delta_{b \rightarrow R} = \sqrt{\gamma_0 d_{B \rightarrow R} ^{- \eta_{bR}}}  \\
&\Delta_{R \rightarrow i} = \sqrt{\gamma_0 d_{R \rightarrow i} ^{- \eta_{Ri}}}  \\ 
&\Delta_{b \rightarrow i} = \sqrt{\gamma_0 d_{B \rightarrow i} ^{- \eta_{bi}}}
\end{align}}
where $\gamma_0$ is the path-loss average channel power gain at a reference distance $d_0 =$1m, $\eta_{k}$ is the path-loss exponent for the wireless link $k \in \{bR,Ri,bi\}$, $d_{R \rightarrow i}$ represents the distance between the RIS and $i$th destination, $d_{B \rightarrow i}$ represents the distance between the BS and $i$th destination,
and $d_{B \rightarrow R}$ represents the distance between the BS and RIS. The small scale fading of the direct links between the BS and the destinations is modelled as a Rayleigh fading channel with zero mean and unit variance \cite{elhattab2021reconfigurable}. Meanwhile, the communication links between the BS and the RIS and between the RIS and the destinations are considered to have LoS components. These links experience small-scale fading that is modelled as Rician fading \cite{elhattab2021reconfigurable}. Accordingly, for all $t \in \mathcal{T}$ and $n \in \mathcal{N}$, the small-scale fading $\hat {\boldsymbol h}_{b \rightarrow R,n}(t)$ between the BS and the RIS on the $n$th frequency resource is defined as:
\begin{equation}
    \hat {\boldsymbol h}_{b \rightarrow R,n}(t) = \sqrt{\frac{K_1}{K_1 + 1}}  \tilde {\boldsymbol h}_{b \rightarrow R,n}(t) + \sqrt{\frac{1}{K_1 + 1}}  \bar {\boldsymbol h}_{b \rightarrow R,n}(t),
    \label{ssf1}
\end{equation}
\noindent
where $K_1$ is the Rician factor, and $\tilde {\boldsymbol h}_{b \rightarrow R,n}(t)$ and $\bar {\boldsymbol h}_{b \rightarrow R,n}(t)$ are the deterministic LoS and Rayleigh fading components respectively. Similarly, for all $t \in \mathcal{T}$, $i \in \mathcal{I}$, and $n \in \mathcal{N}$, the small-scale fading $\hat {\boldsymbol h}_{R \rightarrow i,n}(t)$ between the RIS and the $i$th destination on the $n$th frequency resource is given as:
\begin{equation}
    \hat {\boldsymbol h}_{R \rightarrow i,n}(t) = \sqrt{\frac{K_2}{K_2 + 1}}  \tilde {\boldsymbol h}_{R \rightarrow i,n}(t) + \sqrt{\frac{1}{K_2 + 1}}  \bar {\boldsymbol h}_{R \rightarrow i,n}(t),
    \label{ssf2}
\end{equation}
\noindent
where $K_2$ is the Rician factor and $\tilde {\boldsymbol h}_{R \rightarrow i,n}(t)$ and $\bar {\boldsymbol h}_{R \rightarrow i,n}(t)$ are the deterministic LoS and Rayleigh fading components respectively. Additionally, similar to other works in literature \cite{yang2021reconfigurable,ni2021resource,elhattab2022ris}, we assume that the channel state information (CSI) of the considered wireless links is perfectly estimated at the BS. Although, obtaining the perfect CSI is quite challenging, recent studies \cite{he2019cascaded,wei2021channel} have provided means to obtain efficient channel estimation techniques for RIS-enabled networks that can be embraced with our system model to obtain accurate CSI.    

\par {Based on the above discussion, and for all $t \in \mathcal{T}$, $i \in \mathcal{I}$, and $n \in \mathcal{N}$, the signal-to-noise ratio (SNR) at the $i$th destination at the $t$ time-slot and for the $n$th channel can be expressed as
\begin{equation}
    \gamma_{i,n}(\boldsymbol \Phi(t)) = \frac{P |\boldsymbol h^H_{b \rightarrow R,n}(t) \boldsymbol \Phi(t) \boldsymbol h_{R \rightarrow i,n}(t) + h_{b \rightarrow i,n}(t) |^2} {\sigma^2},
    \label{snrconstraint}
\end{equation}}
\noindent
where $\sigma^2$ is the noise power experienced at each destination and $P$ is the transmit power of the BS. So far, we have discussed the main components related to the SNR at each destination. Next, we will discuss the main elements for the AoI problem.
\section{Age of Information}
\label{aoiDefinition}
{The AoI illustrates how old the information is from a destination's perspective and is defined as the time elapsed since the most recent successful transmission of the valid information update \cite{kosta2017age}}. {For all $t \in \mathcal{T}$ and $i \in \mathcal{I}$, let $y_i(t)$ denote the AoI for a destination $i$ in time-slot $t$}. In addition, it is important to mention that a successful delivery of a packet at the destination in a given time slot $t$, for all $t \in \mathcal{T}$, is conditioned on two realizations:
\begin{enumerate}
    \item The stream selected by the BS for scheduling in time-slot $t$ has a packet available in its queue.
    \item The SNR of the channel between the BS and the destination including the impact of both the direct and indirect links is above a given threshold.
\end{enumerate}
Precisely, for all $i \in \mathcal{I}$ and $t \in \mathcal{T}$, if a packet of the $i$th traffic stream is scheduled by the BS and it is successfully delivered at the $t$th time-slot\footnote{{In this work, we considered that the transmission of each packet occupies one time-slot from BS to each destination \cite{kosta2020non}}.}, then the corresponding AoI in the subsequent time-slot will be given by $ y_i({t+1})= z_i(t) +1$, where $z_i(t)$ represents the system time of the packet in queue $i$ at the beginning of slot $t$. Conversely, if the transmission remained unsuccessful, then the AoI in the subsequent time-slot will be given by $y_i({t+1}) = y_i({t})+1$. Hence, for all $i \in \mathcal{I}$, the evolution of AoI of destination $i$ \cite{kadota2019minimizing} is given as
{
\begin{equation}
  y_i({t+1}) =
 \begin{cases}
  {z_i({t}) +1} & \text{if  } x_{i,n}({t})=1, \, \beta_i({t})=1, \text{and} \\
 &  \gamma_{i,n}(\boldsymbol \Phi(t)) \geq \gamma_{\rm th},
  \\
   y_i({t})+1 &\text{otherwise,}
 \end{cases}
 \label{mainage}
\end{equation}}

\noindent
where $y_i({0}) = 0$ and $\beta_i({t})$ is a binary variable that indicates whether the $i$th stream has an available packet for transmission at the beginning of time-slot $t$ or not. It is worth mentioning that, for all $i \in \mathcal{I}$, the value of $z_i$ is reset to $0$ when a new packet of the $i$th stream arrives in its queue. However, if no new packet is available at the $i$th queue, then the value of $z_i$ is linearly increased by $1$ in the subsequent time-slot. Based on this, for all $i \in \mathcal{I}$, the evolution of $z_i$  \cite{kadota2019minimizing} is given as
\begin{equation}
 z_i(t+1) =
 \begin{cases}
  {0} & \text{if } u_{i}({t+1}) =1 , \forall i,t.   \\
 z_i({t})+1 &\text{otherwise.}
 \end{cases}
 \label{cons1}
\end{equation}
In addition, it is important to mention that, for all $i \in \mathcal{I}$, the value of $\beta_i({t})$ changes to $0$ only when the packet of stream $i$ is scheduled and successfully delivered and there is no new arrival in the same queue, i.e., $u_i(t)=0$. {Based on this, for all $i \in \mathcal{I}$, the evolution of $\beta_i({t})$ \cite{9285215} can be written as:
\begin{equation}
 \beta_i({t+1}) =
 \begin{cases}
  {1} & \text{if  } u_{i}({t+1}) =1,    \\
  {0} & \text{if  } \beta_i({t}) x_{i,n}({t}) =1 \,\,  \wedge  \\
  &   \gamma_{i,n}(\boldsymbol \Phi(t)) \geq \gamma_{\rm th}, \\
 \beta_i({t}), &\text{otherwise.}
 \end{cases}
 \label{consbeta}
\end{equation}
which can be rewritten as
\begin{equation}\label{bett1}
    \beta_i({t+1}) = u_i({t+1}) + \beta_i({t})(1-x_{i,n}({t}))(1-u_i({t+1})). 
\end{equation}}
\noindent

For the sake of tractability, the AoI can be explained by the following  \cite{9285215}:
{\begin{fleqn}
\begin{equation}
     y_i({t+1}) = 1 + x_{i,n}(t)\beta_{i}(t)z_i({t})  + (1-x_{i,n}(t) \beta_{i}(t))y_i({t})\label{age1}  \end{equation}
\end{fleqn}
% \begin{equation}
%      (1-x_{i,n}(t))\beta_{i}(t)y_i({t})+ x_{i,n}(t)(1-\beta_{i}(t))y_i({t}) +1
%      \label{age1}
% \end{equation}
\begin{equation}
    \gamma_{i,n}(\boldsymbol \Phi(t)) \geq x_{i,n}(t)\beta_{i}(t) \gamma_{\rm th},
    \label{age2}
\end{equation}}
%\subsection{Illustrative Example}
\begin{figure}[t]
\centering
\includegraphics[width=1\linewidth]{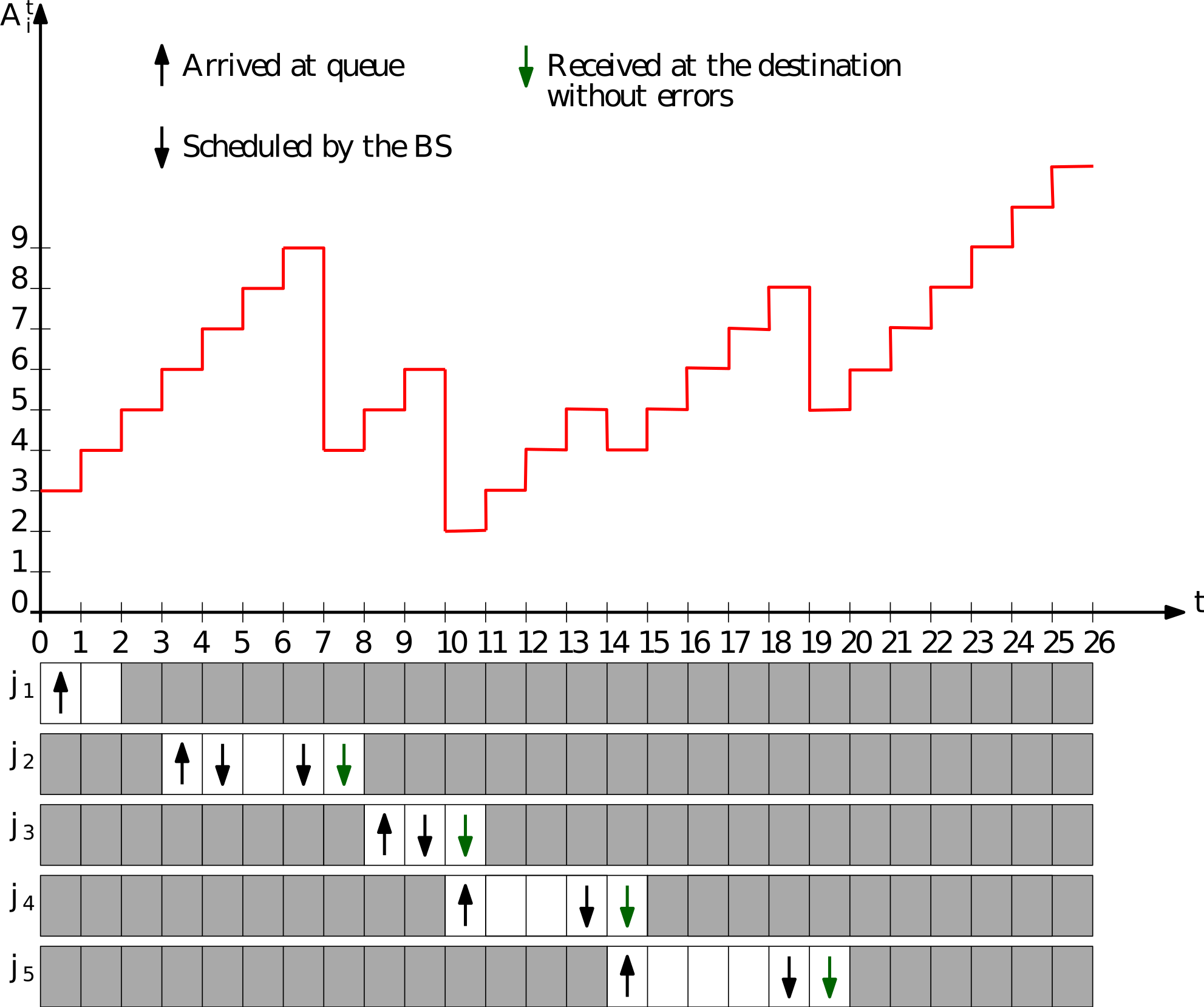}
\caption{An example of evolution of AoI}
\label{aoievolution}
\end{figure} 
To better understand the definition of AoI and to determine its calculation in the studied system model, an example is provided by Fig. \ref{aoievolution}, which  illustrates the evolution of AoI associated with one traffic stream $i \in \mathcal{I}$ over the $T$ time slots. Recall that packets at each queue may arrive at different times and the AoI increases linearly at every time-slot between the two successfully received updates. A packet delivery is considered successful if the received SNR at the $i$th destination is above the threshold, which can be achieved by properly adjusting the phase shifts of the RIS elements. In this example, we consider that five packets of the $i$th stream, indexed from $j_1$ to $j_5$, arrive to the system at different time slots. We assume the system is running for some time already, i.e., $y_i(t)$ at the beginning of time-slot $t=1$ has an initial value of $3$. Assume that the first packet $j_1$ arrives to the queue when there was no other packet in the system. In this case, the value of $z_i(t)=0$ at time-slot $t=1$. At $t=2$, $j_1$ is waiting to get scheduled by the BS and $z_i(t)$ evolves to $1$. Assume that at $t=3$, a fresh packet $j_2$ arrives. The arrival of $j_2$ causes the $j_1$ to get discarded and resets the $z_i(t)$ to 0. However, $y_i(t)$ still increases linearly. At $t=4$, the $j_2$ is scheduled but the delivery remained unsuccessful probably due to the channel conditions. However, another scheduling of $j_2$ at $t=6$ resulted in a successful delivery at $t=7$, which causes the age to drop. Afterwards, at $t=8$, $t=10$ and $t=14$, the packets $j_3$, $j_4$ and $j_5$ are arrived back to back and were scheduled and delivered to the $i$th destination such that the SNR was above the threshold. Thus, the delivery of packets without errors causes the AoI to get reduced. Fig. \ref{aoievolution} demonstrates that the AoI minimization depends not only on the frequent arrivals or the persistent scheduling, but also on the successful delivery at the destination which is challenging due to the wireless channel impairments in the system. Fortunately, the RIS will play a big role in dealing with this issue. In fact, by efficiently configuring the phase shifts of the reflective elements of the RIS, the received signals strengths can be improved at the destination, which increases the chances of the successful delivery of the packets and ultimately helps to reduce the AoI.

\section{Problem Formulation}
\label{formulationprob}
{In this section, we leverage the communication model and the AoI definition presented in the previous sections to formulate a joint optimization of packets scheduling and RIS configuration to minimize the AoI of the system}. 
\subsection{Problem Formulation}
{To ensure the freshness of the received information at each destination, we aim to minimize the expected sum AoI for the $I$ streams over the time horizon of $\mathcal{T}$. Let $\mathcal{X}$ and $\mathcal{R}$ denote the sets of the scheduling policies and the RIS configurations over the time horizon $\mathcal{T}$, which are defined, respectively, as}

{
\begin{align}
&\mathcal{X} = \left\{x_{i,n}(t)|\,\, \forall t \in \mathcal{T}, i \in \mathcal{I}, n \in \mathcal{N} \right\},  \\
&\mathcal{R} = \left\{ \boldsymbol{\Phi}(t)|\,\, \forall t \in \mathcal{T} \right\}.
\end{align}}
{Hence, the optimization problem can be formulated as:
\begin{subequations} 
\begin{align}
\label{obj_m} \mathcal{OP}: \quad & \min_{\mathcal{X}, \mathcal{R}} \dfrac{1}{I}\dfrac{1}{T}\mathbb{E}
 \Big[ \sum_{t=1}^T \sum_{i=1}^ I    y_{i}(t)|y_{i}(0)=0 \Big], \\
& \text{s.t.}  \quad \eqref{cons_channel1}, \eqref{cons_channel2}, \eqref{bett1}-\eqref{age2},  \notag \\
&\qquad\,\, \theta_f(t) \in [0,2\pi), \quad\,\, \forall t \in \mathcal{T},\,\,f \in \mathcal{F}, \label{constheta} \\ 
&\qquad\,\, x_{i,n}(t)  \in \{0,1\}, \quad \forall t \in \mathcal{T},\,\, i \in \mathcal{I},\,\, n \in \mathcal{N} \label{consX},
\end{align}
\end{subequations}
In problem $\mathcal{OP}$, the objective function in \eqref{obj_m} seeks to minimize the expected sum AoI. On the other hand, constraint~\eqref{cons_channel1} ensures that no more than $N$ traffic streams are scheduled for transmission in a given time-slot and constraint~\eqref{cons_channel2} guarantees that each traffic stream is scheduled on at most one frequency channel. Moreover, constraint~\eqref{bett1} shows the current status of the queue of each information stream at each time slot whether it is empty or has a packet available for transmission.
In addition,  constraints~\eqref{age1} and \eqref{age2} ensure the correct evolution of AoI over the time horizon $\mathcal{T}$ considering that the received SNR is above a certain threshold at each time slot. 
Furthermore, constraint~\eqref{constheta} restrains the range of the phase shift at each RIS element. Finally, constraint~\eqref{consX} ensures the binarity of the traffic streams scheduling variables over the available frequency channels at each time slot.
%Finally, constraint~\eqref{constheta} restrains the range of the phase shift at each RIS element. 
Given the uncertainties in the arrival of packets from each traffic stream at a given time-slot, $\mathcal{OP}$ is a stochastic optimization problem over the time horizon $\mathcal{T}$. We further observe that $\mathcal{OP}$ is a mixed-integer non-convex optimization problem which is difficult to be solved. This is due to the existence of both binary decision variables for packet scheduling and the RIS phase shift optimization. Therefore, we solve the $\mathcal{OP}$ by using the concept of bi-level optimization \cite{bard2013practical}.}      
%\par \textcolor{red}{The challenge of the entire approach is how to solve problem OP. Therefore, it is important to provide some description for this problem: 1) stochastic or deterministic?, 2) mixed integer or no?, 3) linear or non linear?, 4) NP hard or no?, 5) is it straightforward to solve it or no?.}
% Generally, solving phase-shift matrix is quite challenging when the objective is to tune the phase shift for multiple traffic streams such that the total channel gain is maximized. The reason being that if the phase-shift matrix is tuned to optimally serve one stream, the other streams might not achieve the required SNR which would subsequently lead to an increase in their AoI.

 \begin{figure*}
\centering
\includegraphics[width=0.8\linewidth]{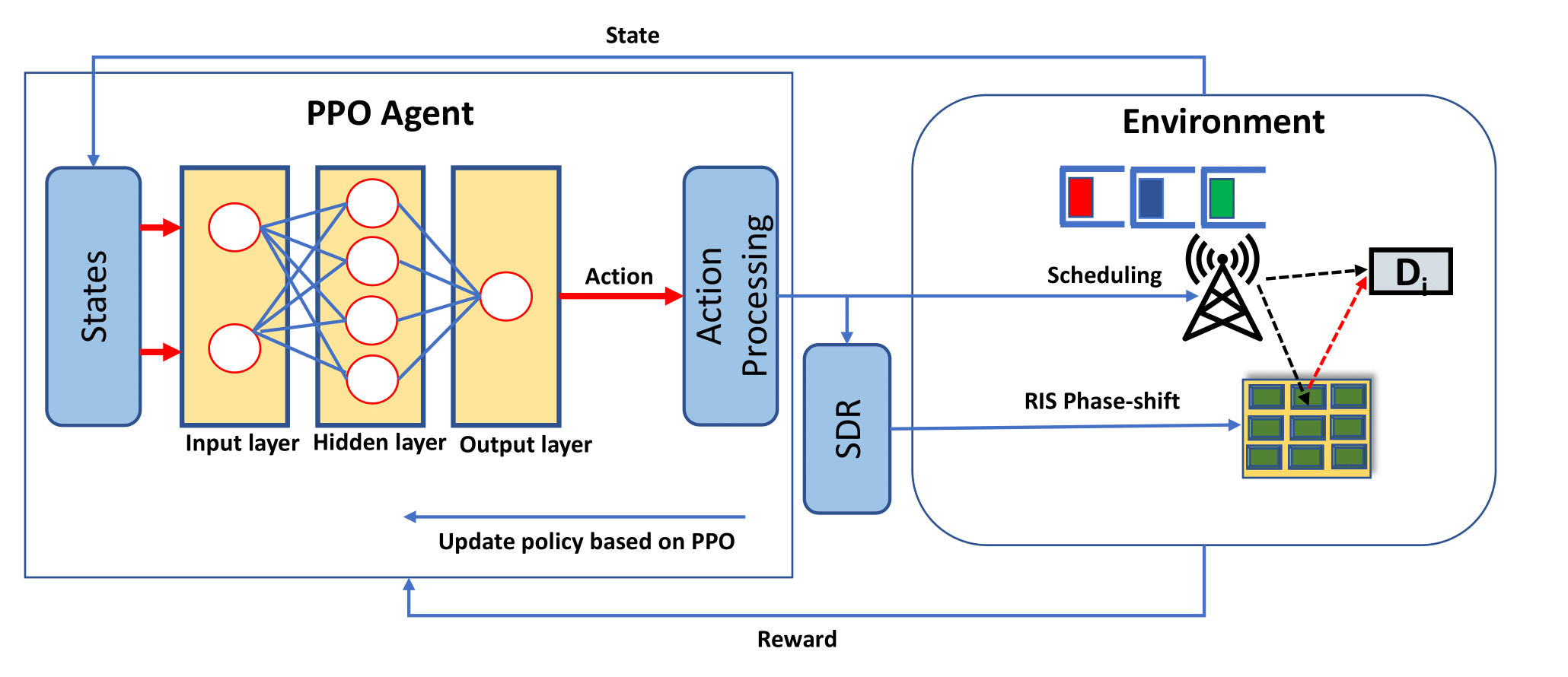}
\caption{An illustration of the proposed solution}
\label{drl}
\end{figure*}

\subsection{Solution Approach}
\label{japsoProblem}
{In this section, we  present our roadmap to solve the joint scheduling and RIS phase shift optimization problem with the objective of minimizing the expected sum AoI. Leveraging the concept of bi-level optimization, we decompose the above problem into an outer traffic stream scheduling problem and an inner  phase shift matrix optimization problem. The stochastic arrival of the traffic into each stream makes the outer problem quite challenging. Hence, we resort to DRL to observe the environment and train an agent that performs scheduling. While, the inner problem of phase shift matrix  optimization is solved using SDR technique. The schematic illustration of our proposed solution approach is presented by Fig. \ref{drl}. We now discuss the two problems in detail: 
\subsubsection{Traffic streams scheduling problem}
The outer problem aims to obtain the traffic stream scheduling having the RIS phase shift matrix obtained from the $\mathcal{OP}_{inner}$ problem is modelled as an MDP. A DRL based on Proximal Policy Optimization (PPO) algorithm is hereby proposed to determine the policy that governs the scheduling of traffic streams. The $\mathcal{OP}_{outer}$ can be written as:

\begin{subequations} 
\begin{align}
\label{obj_m1} \mathcal{OP}_{outer}: \quad & \min_{\mathcal{X}} \dfrac{1}{I}\dfrac{1}{T}\mathbb{E}
 \Big[ \sum_{t=1}^T \sum_{i=1}^ I    y_{i}(t)|y_{i}(0)=0 \Big], \\
& \text{s.t.}  \quad \eqref{cons_channel1}, \eqref{cons_channel2}, \eqref{bett1},\eqref{age1}, \eqref{consX}  \notag \\
&\qquad\,\,  \mathcal{R} = \mathcal{OP}_{Inner} \label{consR}, 
\end{align} 
\end{subequations}
An MDP is generally defined as a 4-tuple ($\boldsymbol S, \boldsymbol A,  \boldsymbol R,\boldsymbol P$), where: $\boldsymbol S$ is a finite set of all possible states $s(t)$ at any time-slot $t$, where  $s(t) \in \boldsymbol S$; $\boldsymbol A$ is a set  of all feasible actions $a(t)$ at any time-slot $t$, where $a(t) \in \boldsymbol A$; $\boldsymbol R$ is the reward distribution, given by a measurable function   
$P(r(t)|s(t), a(t))$, which grants immediate reward $r(t) \in \boldsymbol R $ after an action $a(t) \in \boldsymbol A $ has been chosen in a state  $s(t) \in \boldsymbol S$ at time-slot $t$; $\boldsymbol P$ is a Markovian transition model, where $P(s({t+1})|s(t), a(t)),s(t), s({t+1}) \in \boldsymbol S, a(t) \in \boldsymbol A$ represents the probability of going from state $s(t)$ to state $s({t+1})$ with action $a(t)$. We will next elaborate the state, action and reward functions under the MDP framework as under:
\begin{itemize}
    \item $\textbf{State} \,\,  \mathbb{\boldsymbol  S}$: The system state at time $t$ is defined as $s(t) = (\pmb{y}(t), \boldsymbol \beta(t) , \boldsymbol Z(t)
    )$, where  $s(t) \in \boldsymbol S$. The $\pmb{y}(t)=(y_1(t), y_2(t), ......., y_I(t))$ is a vector of size $I$ containing the AoI of all traffic streams at time-slot $t$,  $\boldsymbol \beta(t)=(\beta_1(t), \beta_2(t), ......., \beta_I(t))$ is a vector of size $I$ containing the indicator that  traffic streams have packets available for transmission and $\boldsymbol Z(t)=(Z_1(t), Z_2(t), ......., Z_I(t))$  is the system time related to the $I$ streams at time slot $t$.
    \item $\textbf{Action} \,\, \mathbb{\boldsymbol A}$ : An action $a(t)$ is executed at each time-slot $t$ denoted by $a(t) \in \boldsymbol A$ consists of channel allocation decisions. The $a(t)$ is a  vector of size $\alpha,$ where $\alpha$ represents the number of channels to be assigned to users. 
    %The total number of actions are computed as $\alpha!$. 
    %Consider a small network instance with 2 channels ($N$)  and 3 users ($\alpha$) on average to be active in each time slot out of 5 total users. The total number of actions given by $U$ is 6 which can easily verified by drawing all possible combinations as (1,2), (2,1), (1,3), (3,1), (2,3), and (3,2), where (1,2) shows that user 1 is scheduled to channel 1 and user 2 is allotted channel 2 and its reciprocal in the case of (2,1).
    \item $\textbf{Reward} \,\, \mathcal{\boldsymbol R}$: The immediate reward $r(t)$ at time slot $t$  is the negative summation of AoI, $r(t) = - \sum_{i=1}^I y_i(t)$, where $r(t) \in \boldsymbol R$.  Considering the objective of minimizing the expected sum AoI, the RL-agent aims to optimize the scheduling decision that leads to minimize the AoI.
     \end{itemize}}
     \noindent
    {
    Algorithm 1 presents our proposed approach with DRL exploiting the PPO to develop the agent. The agent based on PPO is usually implemented in Actor-Critic framework. 
    We now summarize the steps of algorithm. The agent first initializes a random sampling policy and a value function for neural networks as given by (line 3 and line 4). 
    Further, at each episode, the agent observes the environment which is composed of current AoI of all the destinations, the current system time in each queue up to $t$ slot. Then at each time-slot, the agent selects an action which is a vector carrying the channels in a specific order to be mapped with the traffic streams that have a packet available for transmission. The action results to invoke the SDR (Algorithm 2) in order to configure the RIS phases shift matrix to maximize the channel gain. Eventually, the time step reward is calculated which is the negative sum of age of information of all the streams. Once the set of samples have been gathered and rewards have been computed, the agent determines the advantage function (line 15) which is the resultant of the difference of the expected value function from the actual reward. This is to note that the advantage estimate helps the system to analyze how good it is performing based on its normal estimate function value. Regarding the complexity of Based on \cite{9285215}, the total computational complexity 
    of DRL frameworks such as PPO algorithm can be expressed as the number of multiplications: $O(\sum_{p=1}^{P-1} n_p.n_{p-1})$, where $n_p$ is the number of neural units in the $p$-th hidden layer.}
    %The complexity of connected networks with $P$ layers is $O(\sum_{p=0}^P n_p n_{p-1})$ ) where $n_p$ is the total number of neurons in layer $p$ \cite{shokry2020leveraging}. }  
\begin{algorithm}[t]
%   \begin{algocolor}
%\SetAlgoNoLine
\caption{Proposed solution approach for minimizing the expected sum AoI}
\label{algo1}
\textbf{Input:}
 Number of users~($I$), Number of time-slots~($T$), Learning Rate, Episodes $K$, threshold ($\gamma_{\rm th}$).\\
\textbf{Output:} User scheduling, Resource allocation and Phase shift matrix.\\
Initialize policy $\pi$ with random parameter $\theta$  \\
Initial value function $V$ with random parameters $\phi$ \\
\For{$k \leftarrow 1: K$}{
\For{$t \leftarrow 1: T$}{
Get $(\pmb{y}(t), \boldsymbol \beta(t), \boldsymbol Z(t)
    )$ from the environment.  \\
sample action $a(t) \sim $ $\pi_{\theta_{old}}.$ \\
Take action $a(t)$ that specifies the channels (in a specific order).\\
Obtain the resource allocation by mapping the top $N$ traffic streams that have a packet available for transmission.\\
Configure $\boldsymbol \Phi(t)$ that maximizes the SNR of the mapped users to the respective channels  using SDR approach using Algorithm 2. \\
Perform the feasibility check to determine if SNR threshold constraint is satisfied.\\
Get relevant reward $r(t)$ and $s(t+1)$. \\
Store $(s(t), a(t), r(t), s(t+1))$ as one transition in the experience replay.
}
Compute advantage estimate $\hat A$ for all epochs. \\
Optimize surrogate loss function using Adam optimizer. \\
Update current policy $\pi_{\theta_{old}} \leftarrow \pi_{\theta}.$ \\
%$A_i(1) \leftarrow$ 1
}
% \end{algocolor}
\end{algorithm} 
\subsubsection{SDR for RIS phase shift coefficients}
\begin{algorithm}[t]
\caption{Design of Phase Shift Matrix via SDR}
\label{algo2}
\textbf{Input:} Number of users, Number of RIS elements\\
\textbf{Output:}  Phase shift matrix, i.e. $\boldsymbol{\Phi}$.  \\
Initialize the maximum generation of candidate random vector as $\xi$ \\
Solve the relaxed SDR problem \eqref{objop2}. \\
\If{rank($\boldsymbol{V}$) = 1}
{With the obtained $\boldsymbol{V}$, calculate the eigenvalue $\omega$ and eigen vector $\boldsymbol u$ according to $\boldsymbol Vu = \omega u$.\\
Update the value of the phase-shift matrix $\boldsymbol{\Phi^*}:= diag(\sqrt{\omega \boldsymbol u})$.}
\Else{obtain the eigenvalue decomposition using  Eq. \eqref{neweqq} \\
\For{$x \leftarrow 1: \xi$}{Generate a Gaussian random vector $\boldsymbol r_x,$ i.e., $\boldsymbol{r_x} \sim CN(0; I_{F+1})$  \\
Obtain a candidate solution $\boldsymbol{\Theta_{x}}$ using Eq. \eqref{eqnew1} and Eq. \eqref{neweq3}.}
Find the optimal  $\boldsymbol{\Theta^*}:= \boldsymbol{\Theta_{x}}$ that maximizes the combined channel gain for all users.}
\end{algorithm} 
%}
Referring to the definition of AoI given in Section \eqref{aoiDefinition}, if no successful status update is delivered, the age for a destination will increase linearly with the time axis. Therefore, if the updated packets of a stream are scheduled by the BS but the corresponding channels do not satisfy the SNR constraints, the total AoI in $T$ time-slots will increase. Therefore, the phase shifts of the reflective elements should be configured to maximize the SNR of the channels corresponding to the selected streams. The SDR technique is applied to obtain $\boldsymbol{\theta}$ that can maximize the overall channel gain.
{
\begin{subequations} \label{eq:subeqf nscdglobalEQ2}
\begin{align}
\label{objop2}
\text{$\mathcal{OP}_{inner}   $: } & \max_{\theta} \,\,\, |\boldsymbol h^H_{b \rightarrow R,n}(t) \boldsymbol \Phi(t) \boldsymbol h_{R \rightarrow i,n}(t) + h_{b \rightarrow i,n}(t)|^2 \ \\
&
\text{s.t.  \,\,\,  }  \notag \\  & \label{thetaf} 0 \leq \text{$\theta_f(t) \leq 2\pi$, \,\,\, $\forall f \in \llbracket 1,F\rrbracket $} 
\end{align}
\end{subequations}}
% \textcolor{blue}{
% \begin{subequations} \label{eq:subeqf nscdglobalEQ2}
% \begin{align}
% \label{objop2}
% \text{$\mathcal {P} $: } & \text{Find} \,\,\, \boldsymbol \theta \\
% &
% \text{s.t.  \,\,\,  }  \notag \\  & \label{thetaf} 0 \leq \text{$\theta_f(t) \leq 2\pi$, \,\,\, $\forall f \in \llbracket 1,F\rrbracket $} 
% \end{align}
% \end{subequations}}
Let us define, $\boldsymbol v =[ v_1,v_2,....,v_F]^H$, where $v_f=e^{j\theta_f}$, $\forall f$. Thus, the constraints  in \eqref{thetaf} are equivalent to the unit-modulus constraints, i.e., $|v_f|^2=$1 $\forall f \in F$. By applying the change of variables, $\boldsymbol h^H_{b \rightarrow R,n}(t)\boldsymbol  \Phi(t) \boldsymbol  h_{R \rightarrow i,n}(t)$ can be represented as $\boldsymbol v^H  \boldsymbol{\mathcal{ W}}(t)$, where 
$\boldsymbol{\mathcal{W}}(t) =\mathrm{diag}(\boldsymbol h^H_{b \rightarrow R,n}(t))\boldsymbol h_{R \rightarrow i,n}(t)$. Thus, we have
\begin{equation*}
 |\boldsymbol h_{b \rightarrow R,n}(t) \boldsymbol \Phi(t) \boldsymbol h_{R \rightarrow i,n}(t) + h_{b \rightarrow i,n}(t)|^2  
\end{equation*}
\begin{equation}
= | \boldsymbol v^H \boldsymbol{\mathcal{W}}(t) + h_{b \rightarrow i,n}(t)|^2   
\end{equation}
An expression of overall channel gain denoted by $\mathcal{Z}$ can be given as:
\begin{align}
    \mathcal{Z} &= {|\boldsymbol v^H \boldsymbol{\mathcal{W}}(t) + h_{b \rightarrow i,n}(t)|^2}, \cr
    &= \boldsymbol v^H  \boldsymbol{\mathcal{W}}(t)\boldsymbol{\mathcal{W}}^H(t)\boldsymbol v + h_{b \rightarrow i,n}(t)  \boldsymbol{\mathcal{W}}^H(t)\boldsymbol v   \cr &+  {\boldsymbol v^H  \boldsymbol{\mathcal{W}}(t)h_{b \rightarrow i,n}(t) + |h_{b \rightarrow i,n}(t)|^2 },
\end{align}
\noindent
The above equation can be written as follows
\begin{equation}
    \mathcal{Z} = \bar {\boldsymbol v}^H  \boldsymbol{\Phi} \bar {\boldsymbol v} + |h_{b \rightarrow i,n}(t)|^2,
\end{equation}

where  
\[
   \boldsymbol{\Phi}=
    \begin{bmatrix}
   {\begin{array}{cc}
  \boldsymbol{\mathcal{W}}(t)\boldsymbol{\mathcal{W}}^H(t) & \boldsymbol{\mathcal{W}}(t)h_{b \rightarrow i,n}(t) \\
   h_{b \rightarrow i,n}(t)\boldsymbol{\mathcal{W}}^H(t) & 0 \\
  \end{array} } 
  \end{bmatrix}
\],

\[
 \bar {\boldsymbol v}=
  \begin{bmatrix}
   \boldsymbol v  \\
    1  
  \end{bmatrix}
\]
\noindent Note that $\bar {\boldsymbol v}^H  \boldsymbol{\Phi} \bar {\boldsymbol
v}=$ tr$(\boldsymbol{\Phi} \bar {\boldsymbol v} \bar {\boldsymbol v}^H)$. Additionally, we define  $\boldsymbol{V} ={\boldsymbol v} \bar {\boldsymbol v}^H$, which needs to satisfy rank$(\boldsymbol V)$=1 and $\boldsymbol V \geq$ 0. This rank constraint (rank$(\boldsymbol V)$=1) is non-convex \cite{8811733}. By dropping this constraint, the problem $\mathcal {OP}_{inner} $ can be rewritten as: 
{
\begin{subequations} \label{eq:subeqf nscdglobalEQ2}
\begin{align}
\label{objop2}
\text{$\mathcal {P} $1: } & \max_\Phi  \,\,\, \mathcal{Z}(\Phi) \\
& \text{s.t.  \,\,\,  }  \notag \\  & \label{vf} \text{$\boldsymbol{V} \geq$ 0},
  \\
 &\label{vff} [\boldsymbol{V}]_{F,F} = 1.
%   \\ &  \label{vfff}
%   \mathrm{tr}(\boldsymbol{\Theta} \boldsymbol{V}) + |h_{b \rightarrow i,n}(t)|^2  \geq x_{i,n}(t) \gamma_{\rm th} {\sigma^2}, 
\end{align}
\end{subequations}}

After the proposed transformation, the above problem can be solved by any convex optimization solver such as CVX\cite{8811733}. Generally, the optimal $\boldsymbol V$ obtained by solving  problem $\mathcal {P} $1 does not satisfy the rank one constraint. This implies that the optimal solution of the $\mathcal {P} $1 only serves as an upper bound for the problem $\mathcal {OP}_{inner} $. Therefore, other steps are needed to construct a rank one solution. The rank one solution is hence achieved by applying
the Gaussian randomization scheme. 
We now describe it in detail.
Firstly, we obtain the eigenvalue decomposition of $\boldsymbol V$ as
\begin{equation}
\label{neweqq}
 \boldsymbol {V} = \boldsymbol{U} \boldsymbol{\Sigma} \boldsymbol{U}^H,   
\end{equation}
where $\boldsymbol{U} =[ u_1,u_2,....,u_{F+1}]$ is a unitary matrix and $\boldsymbol{\Sigma}$=~ diag($\omega_1, \omega_3,.....,\omega_{F+1}$) is a diagonal matrix, respectively. Next, a
random vector is generated as follows,

\begin{equation}
\label{eqnew1}
\bar {\boldsymbol{v}}=\boldsymbol{U}\boldsymbol{\Sigma}^{1/2}\boldsymbol{r}, \end{equation}
\noindent
where $\boldsymbol r$ is a random vector that follows a circularly symmetric complex Gaussian (CSCG) distribution with a zero mean and a co-variance matrix equal
to the identity matrix of order $F + 1$, denoted by $I_{F+1}$ i.e., $\boldsymbol{r} \sim CN(0; I_{F+1})$ . Furthermore, we
generate the scalar $\boldsymbol{v}$

\begin{equation}
\label{neweq3}
    \boldsymbol{v} = \exp \left[j \, \,  \arg    (\frac{[\bar {\boldsymbol{v}}]_{1:F} } {[\bar {\boldsymbol{v}}]_{F+1} })  \right],
\end{equation}
\noindent
where $[\bar{\boldsymbol{v}}]_{1:F}$ represents the vector with first $F$ elements in $\boldsymbol v$.
It is significant to highlight that the SDR approach followed by a large number of Gauss randomization can guarantee a minimum accuracy of $\pi$/4 of the optimal objective value \cite{8811733}. The core details of the phase shift matrix optimization is given by Algorithm 2. Regarding the complexity of Algorithm 2, obtaining the phase-shift matrix is a semi-definite programming (SDP) problem which can be solved by the interior point method and its order of computational complexity with $m$ SDP constraints that contain an $n \times n$ positive semi-definite matrix is given as $\mathcal{O}(\sqrt{n}\text{log}(1/\epsilon)(mn^3 + m^2n^2 + m^3))$, where $\epsilon>$0 is the solution accuracy \cite{elhattab2021reconfigurable}. The approximate computational complexity to solve SDP can be written as $\mathcal{O }(\text{log}(1/\epsilon)(F^{4.5})$ with $m$ = $F$ and $n$ = $F+1$.
Meanwhile, let $w$ be the maximal number of generated Gaussian random vectors and $T_{GR}$ is the complexity of performing one Gaussian random iteration. Hence, the approximate complexity of obtaining phase shift matrix can be written as $\mathcal{O }(\text{log}(1/\epsilon)(F^{4.5}+ wT_{GR}))$.

 \section{Simulation and Numerical Analysis}
 \label{simulationresults}
 In this section, we present a series of simulations to evaluate the performance of the proposed algorithm. The  simulation parameters are first presented, followed by the adopted benchmark schemes and then the results and discussions. 
 \subsection{Simulations setup}
  \begin{figure}[t]
\centering
\includegraphics[width=1\linewidth]{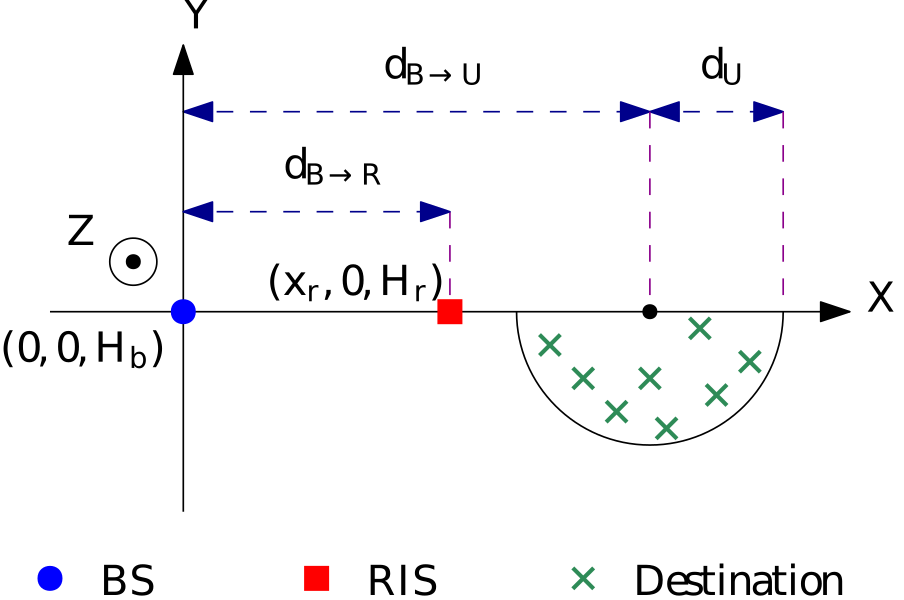}
\caption{Distance Model.}
\label{fig:simulation_setup}
\end{figure}
{We consider a 3-D area where a BS is communicating with a set of spatially dispersed destinations through an RIS. We assume that the global coordinate system $(X, Y, Z)$ is Cartesian. As shown in Fig.~\ref{fig:simulation_setup}, the BS is located at $(0,0,H_{\rm b})$ and the RIS is located at $(x_{\rm r},0,H_{\rm r})$, where $x_{\rm r} = d_{\rm B,R}$ is the distance from the BS to the RIS, and $H_{\rm b}$ and $H_{\rm r}$ are the heights of the transmit antenna of the BS and of the RIS, respectively. In addition, multiple destinations are randomly distributed at the ground level within a given area in the network, where for all $i \in \mathcal{I}$, the locations of destinations are $(x_i,y_i,0)$. Precisely, based on Fig.~\ref{fig:simulation_setup}, the coordinates of the $i$th destination, for all $i \in \mathcal{I}$, are given by
\begin{equation}
    \begin{split}
        x_i = d_{\rm U}\cos(\theta_i) + d_{\rm B,U},\\
        y_i = d_{\rm U}\sin(\theta_i) + d_{\rm B,U},
    \end{split}
\end{equation}
where $d_{\rm U}$ is the radius of the area where the destinations are located, $d_{\rm B,U}$ is the distance from the BS to the center of this area and, $\theta_i \in [\pi, 2 \pi]$ is a polar angle. Unless otherwise indicated, all the simulation parameters are given by Table \ref{tab:simulationParameters}.} 
\begin{table}[t]
    \centering
        \caption{Simulation Parameters.}
        \begin{tabular}{|c|c|}
            \hline
            \textbf{Parameter} &  \textbf{Values}  \\ \hline
            Total number of time slots, T & 100 \\ \hline
            Arrival rate, $\lambda$ & 0.5 \\ \hline
            Activation functions & Softmax and Tanh \\ \hline
            Number of Neurons & 64 \\ \hline
            Number of Hidden layers for Networks & 3 \\ \hline
            Learning Rate & 0.002 \\ \hline
            $d_{\rm B,U}$ & $200$ m \\ \hline
            $d_{\rm U}$ & $10$ m \\ \hline
            $d_{\rm B,R}$ & $200$ m \\ \hline
            $H_{\rm b}$ & $10$ m \\ \hline
            $H_{\rm r}$ & $10$ m \\ \hline
            $\sigma^2$  &  -110 dBm  \\ \hline
            $\eta_{bR}$     & -2.2 \\ \hline
            $\eta_{Ri}$                & -2.2 \\ \hline
            $\eta_{bi}$                 &  -3.5 \\ \hline
            Rician factors (K1, K2) & 2 dB \\ \hline
            Optimizer technique & Adam \\ \hline
            Clip function, $\epsilon$ & 0.2 \\ \hline
            Total number of Epochs & 3000 \\ \hline
            $\gamma_0$ & -20dB \\ \hline
            $\gamma_{th}$ & 45dB \\ \hline
      \end{tabular}
    \label{tab:simulationParameters}
\end{table}
{
\subsection{Benchmark schemes}
To the best of our knowledge, there is no existing approach that aims to solve the problem of minimizing the age of information in RIS-assisted wireless networks by optimizing the scheduling of existing traffic streams and the design of the RIS configuration considering the impacts of the stochastic arrivals of the packets and the multi-user scheduling. Thus, for the sake of comparison, we develop three other baseline schemes in order to assess the performance of the proposed scheme.
\subsubsection{Greedy scheduling with SDR (GS-SDR)}
In this scheme, the scheduling problem is solved using a greedy approach, whereas the RIS configuration problem is solved using the SDR approach. The greedy scheduling approach is explained as follows. At each time-slot $t \in \mathcal{T}$, the traffic streams are first ranked based on their current AoI. The top $N$ streams are selected to get scheduled and the  RIS phase shift matrix optimization is performed to maximize the SNR of these selected streams. If the obtained SNR satisfies the given threshold, the selected streams are assumed to be scheduled and the corresponding age is calculated accordingly. However, the scheduling decisions are taken irrespective of the knowledge that the queue of the selected streams are empty or have packets to deliver. In case, if there is no status update packet in the selected stream's queue, a time-slot is lost.
\subsubsection{Round-Robin scheduling with SDR (RRS-SDR)}
This algorithm is based on round-robin scheme, where at each time-slot, the BS alternately selects an input stream $i \in \mathcal{I}$, starting from the first stream, to upload its status update packet to the destination node. The RIS configuration optimization is performed to maximize the channel gain of the scheduled streams. However, and similar to the GS-SDR baseline, the scheduling decisions are taken irrespective of the knowledge that the queue of the selected streams are empty or have packets to deliver
\subsubsection{DRL with Random Phase-Shift Matrix (DRL-RPM)}
In this approach, the proposed DRL algorithm is used to obtain the scheduling of the traffic streams. However, the RIS configuration is not optimally designed. Instead, a random RIS phase shift matrix is employed.}
\subsection{Results and Discussions}
\begin{figure}[t]
\centering
\includegraphics[width=1\linewidth]{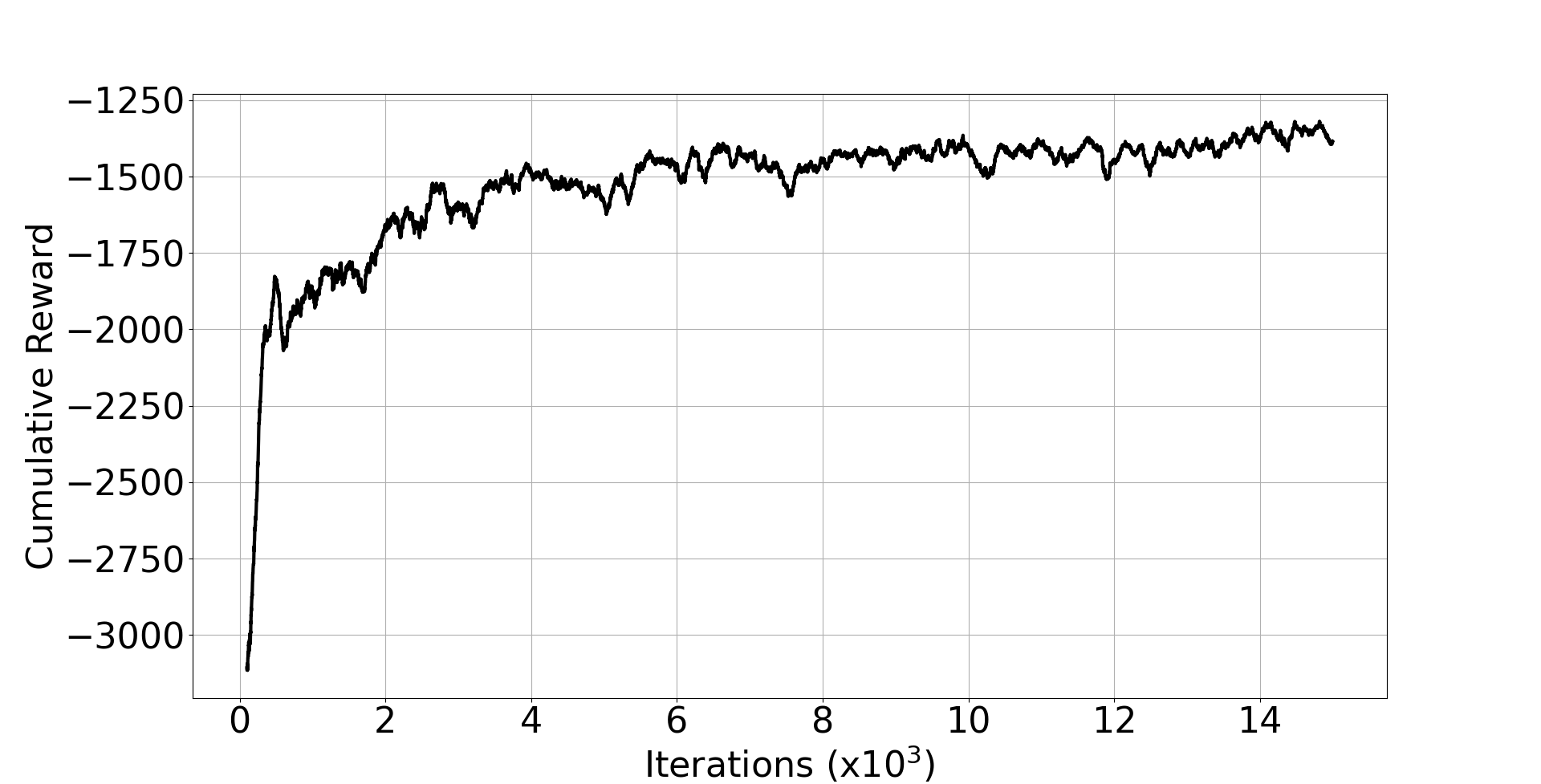}
\caption{Accumulated reward Vs iterations}
\label{convergence}
\end{figure}

{We first attempt to observe the behavior of the DRL agent and to verify the convergence of the proposed algorithm. As depicted in Fig.~\ref{convergence}, the cumulative reward, which is the opposite value of the minimum average sum AoI, is significantly improving as the number of iterations, or episodes, is increasing. Basically, it can be observed from this figure that the proposed PPO algorithm starts to converge after 3000 iterations.}
\begin{figure}[t]
\centering
\subfigure[] {\centering\includegraphics[width=0.5\textwidth]{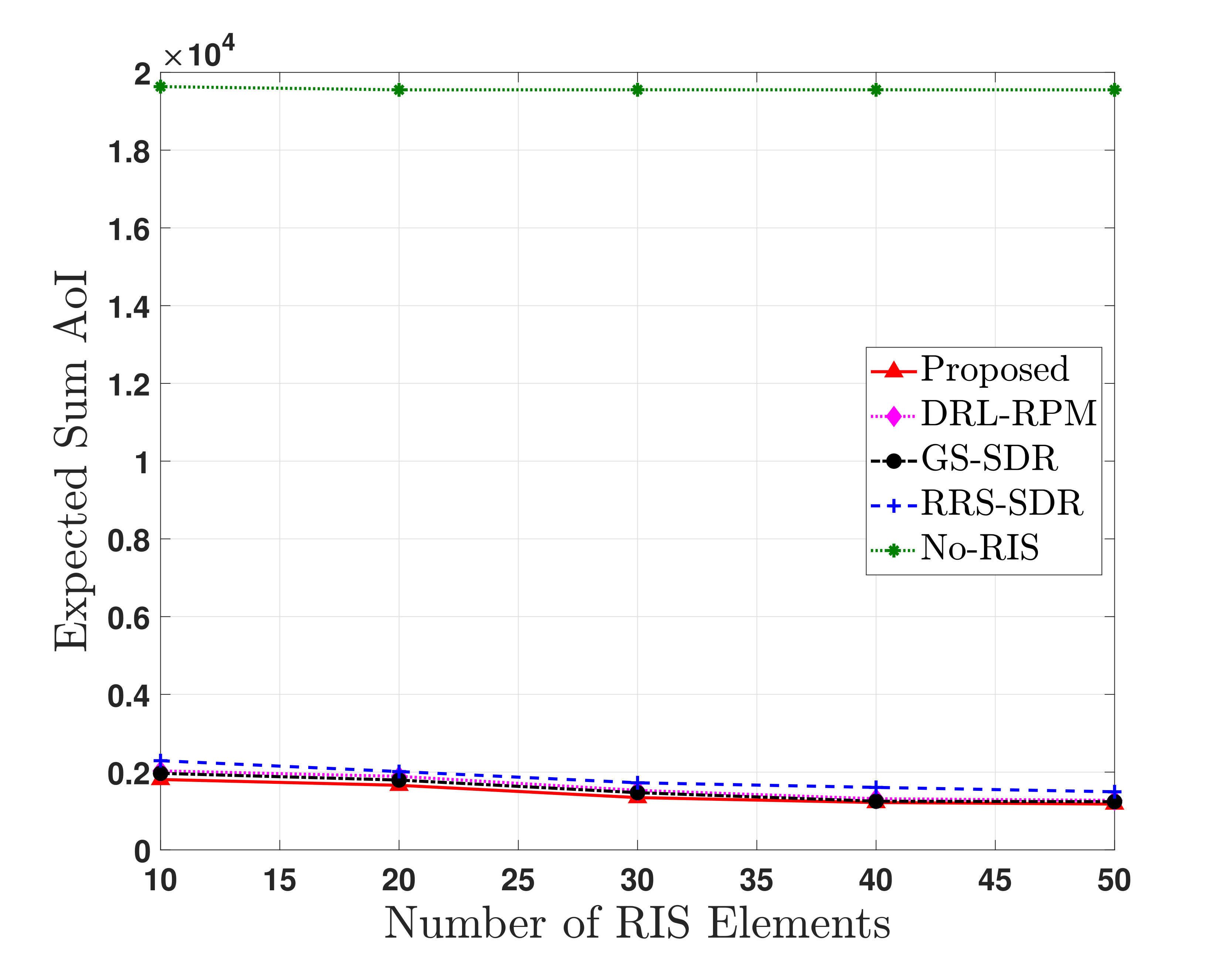}\label{ageVsRISelements_a}} 
\subfigure[Zoomed view of (a)] {\centering\includegraphics[width=0.5\textwidth]{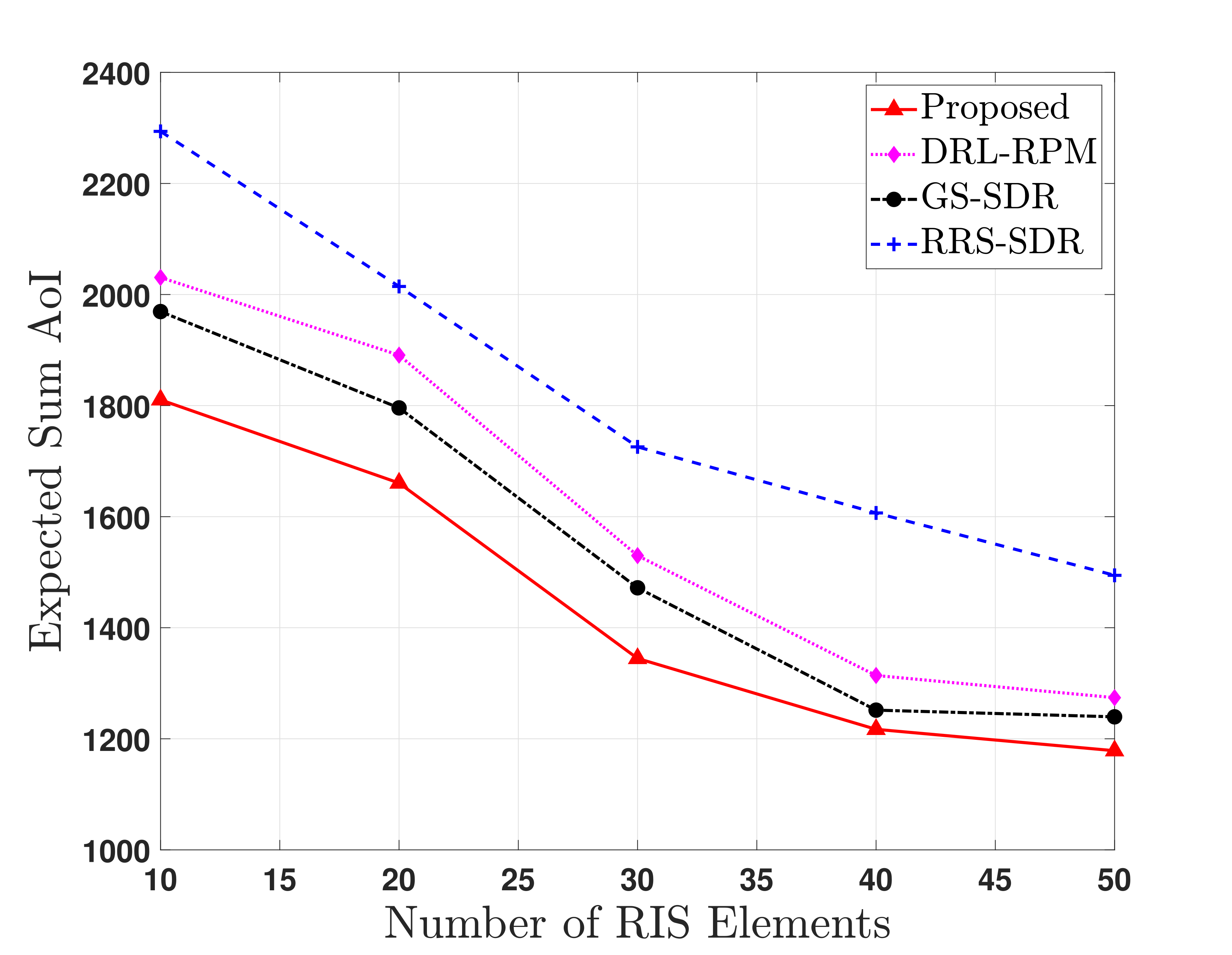}\label{ageVsRISelements_b}}
\caption{Impact of number of RIS elements on the AoI}
\label{ageVsRISelements}
\end{figure}
\begin{figure}[t]
\centering
\includegraphics[width=1\linewidth]{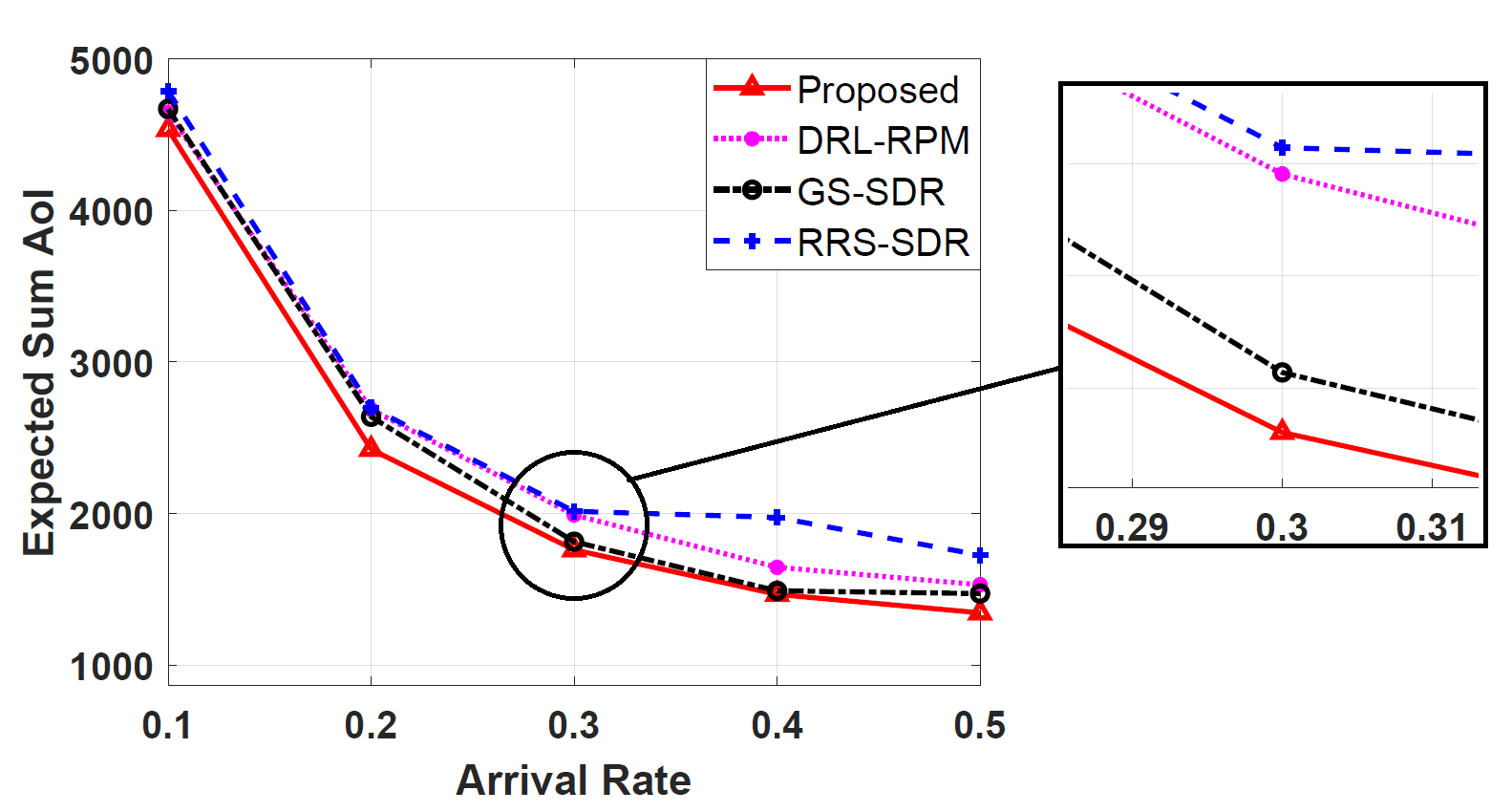}
\caption{Impact of arrival rate on the AoI}
\label{agevsarrivals}
\end{figure} 
{In the next experiment, the impact of varying the size of the RIS (number of elements) on the expected sum AoI under the different schemes is analyzed as shown in Fig.~\ref{ageVsRISelements}. The impact of RIS elements is simulated by varying the number of RIS elements from $10$ to $50$ with a step size of $10$. It can be seen from Fig.~\ref{ageVsRISelements_a} that the integration of the RIS has a significant impact on the AoI as compared to the case when RIS is not utilized, i.e., when the direct links from the BS to destinations are solely relied on to transmit time-sensitive information. Indeed, this shows that the RIS can significantly improve the channel quality of the scheduled users, which subsequently results in a high success rate of packets delivery. We observed that the curves of the  expected sum age of information for all schemes decrease as the number of RIS elements increase. Obviously, the channel quality of the potential scheduled users can be greatly enhanced by increasing the number of RIS elements as it improves the chances of successful delivery at the destination and eventually end up decreasing the AoI. From Fig.~\ref{ageVsRISelements_b}, one can remark that the proposed algorithm achieves the lowest expected sum AoI compared to other benchmark approaches. For example, when the RIS elements is $50$, the expected sum AoI obtained by the proposed algorithm is around $22\%$ lower than the one obtained by the RRS-SDR approach. This is due to the fact that the proposed PPO-based approach leverages the learning of the packet arrivals of the traffic streams and adjusts the RIS configuration accordingly for the streams that have packets to transmit. However, the other approaches do not consider this important factor which eventually results in worse age performance.} 
\par {Although the expected sum AoI of the proposed algorithm is decreased by around $35\%$ when the number of RIS elements are increased from $10$ to $50$ elements, one can further note from Fig.~\ref{ageVsRISelements} that the decrease in the AoI is not linear with the number of RIS elements, where the decrease in the AoI is not sharp when the number of elements are increased from $40$ to $50$, which is $3\%$ in this case.
This can be explained as increasing the number of RIS elements helps to improve the channel gains which eventually leads to satisfy the SNR threshold constraint. However, once it is satisfied, increasing number of RIS elements may not further bring the AoI down. We also observe that the GS-SDR scheme performs better than all other approaches except the proposed approach. Indeed, the greedy approach opts to schedule the streams with the worst AoI by ranking the streams with their AoI. However, since the scheduling decisions are taken irrespective of whether the scheduled stream has a packet available for transmission or not, a waste of resources occurs, which lowers the efficiency of the method. Unlike that, our method learns the presence of packets for scheduling and it is able to attain better performance through more informed scheduling decisions.} 
\par { We next analyze the impact of a variable network load on the AoI, which is depicted in Fig. \ref{agevsarrivals}, where the impact of increasing the network load on the AoI is investigated. The impact of increasing the load is simulated by varying the arrival rate of the packets from $0.1$ to $0.5$, with a step size of $0.1$. The results are plotted for the expected sum AoI versus the arrival rate. The time-horizon used for this experiment is $T = 100$ time slots. As learnt from the theory of AoI, frequent information updates along with their successful delivery results in keeping the information fresh at a destination. Precisely, a low arrival rate leads to an increase in the expected AoI. However, as the arrival rate increases, more fresh packets arrive to the system and  replace the old ones. Hence, under proper propagation environment through the RIS and  a proper packets scheduling, the AoI decreases when the packets arrival rate increases. These facts are validated by Fig. \ref{agevsarrivals}, where we observe that the curves of the expected AoI for all the schemes decrease as the arrival rate increases. On top of this, our proposed method achieves the lowest AoI as compared to the other methods. For example when the arrival rate is increased from  $0.1$ to $0.5$, the  expected sum age is decreased by $70\%$ for the proposed method. We also observe that the GS-SDR scheme performs better than the RR-SDR and the DRL-RPM schemes even when the arrival rate is low.  The reason is related to its scheduling policy and RIS phase shift matrix optimization approach, since the GS-SDR scheme aims to schedule the streams that give the largest decrease in the sum AoI, and hence, results overall in a lower age than the other baseline approaches. } 
\begin{figure}[t]
\centering
\includegraphics[width=1\linewidth]{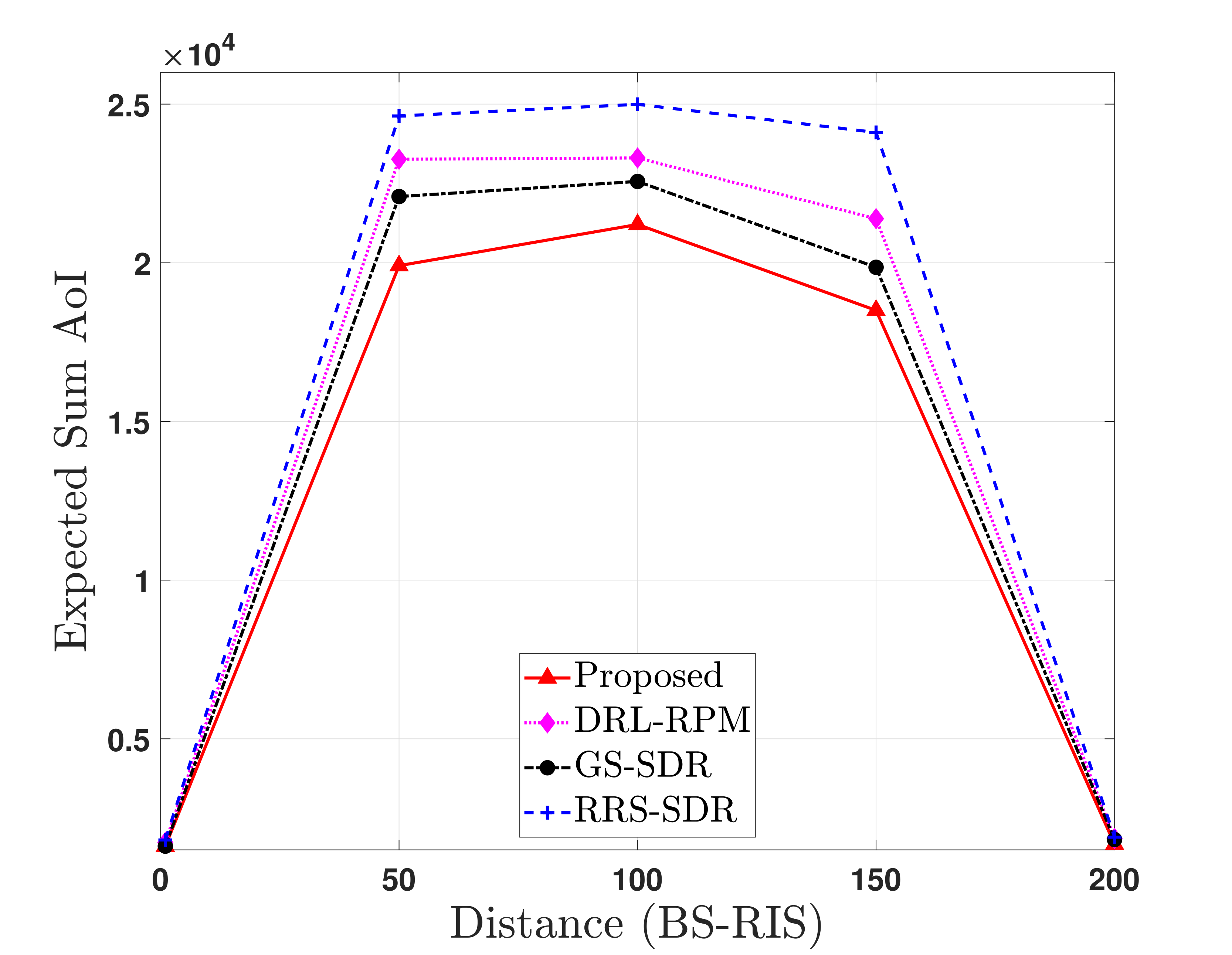}
\caption{Impact of the position of RIS on the AoI}
\label{AgeVsRISdistance}
\end{figure} 
\begin{figure*}[t]
\centering
\subfigure[Proposed] {\centering\includegraphics[width=0.3\textwidth]{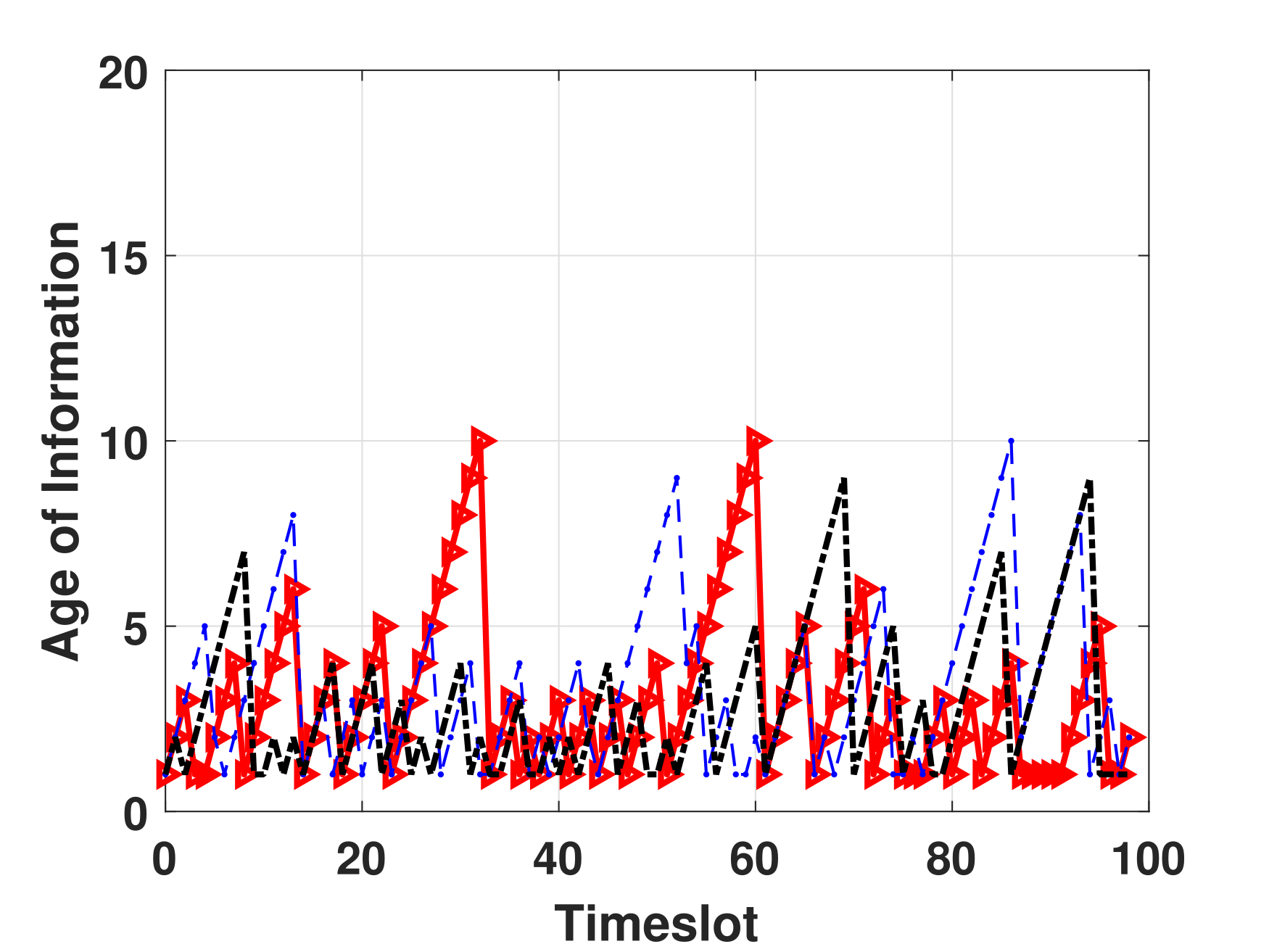}} 
\subfigure[GS-SDR] {\centering\includegraphics[width=0.3\textwidth]{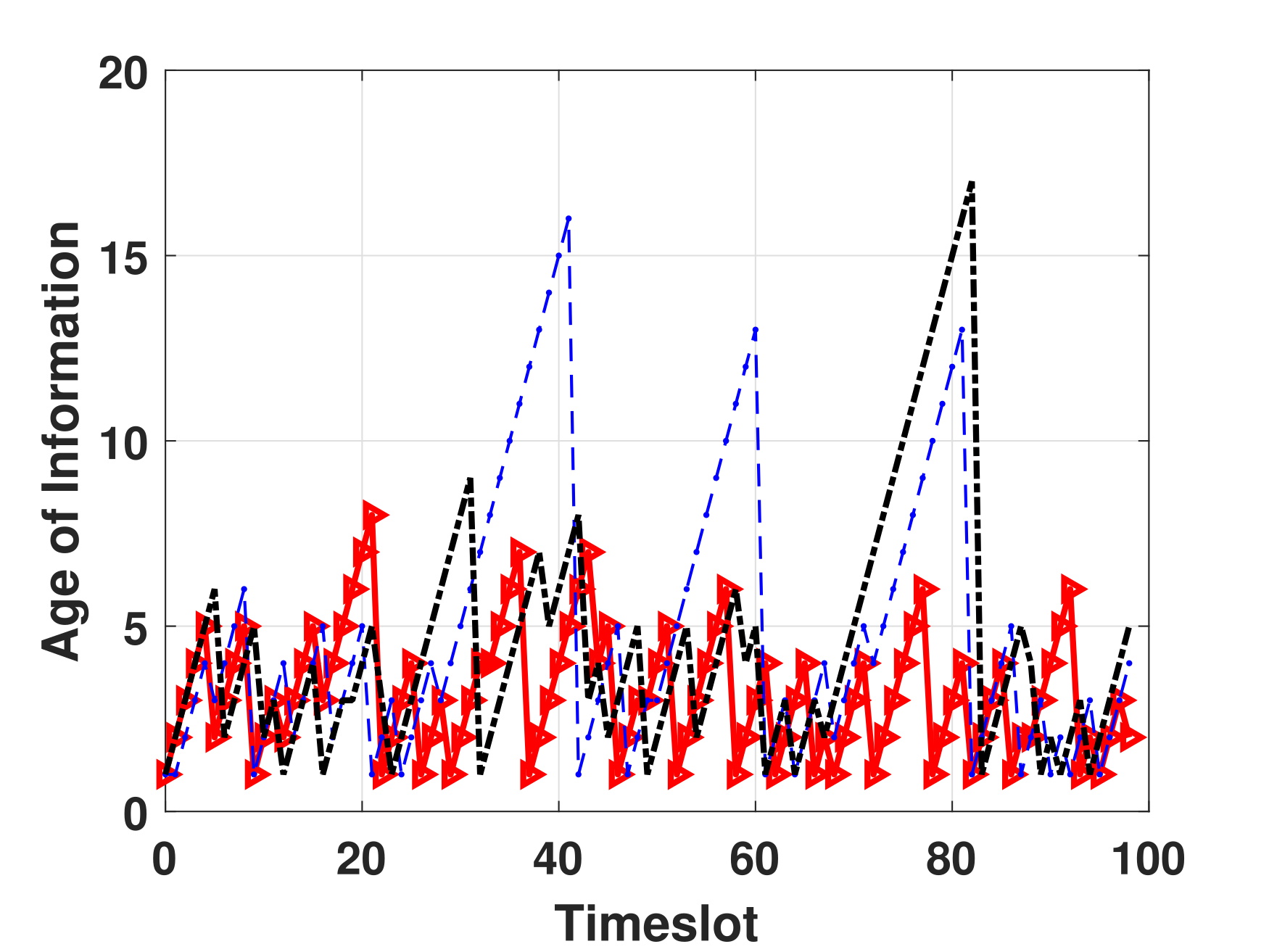}}   \subfigure[RRS-SDR]
{\centering\includegraphics[width=0.3\textwidth]{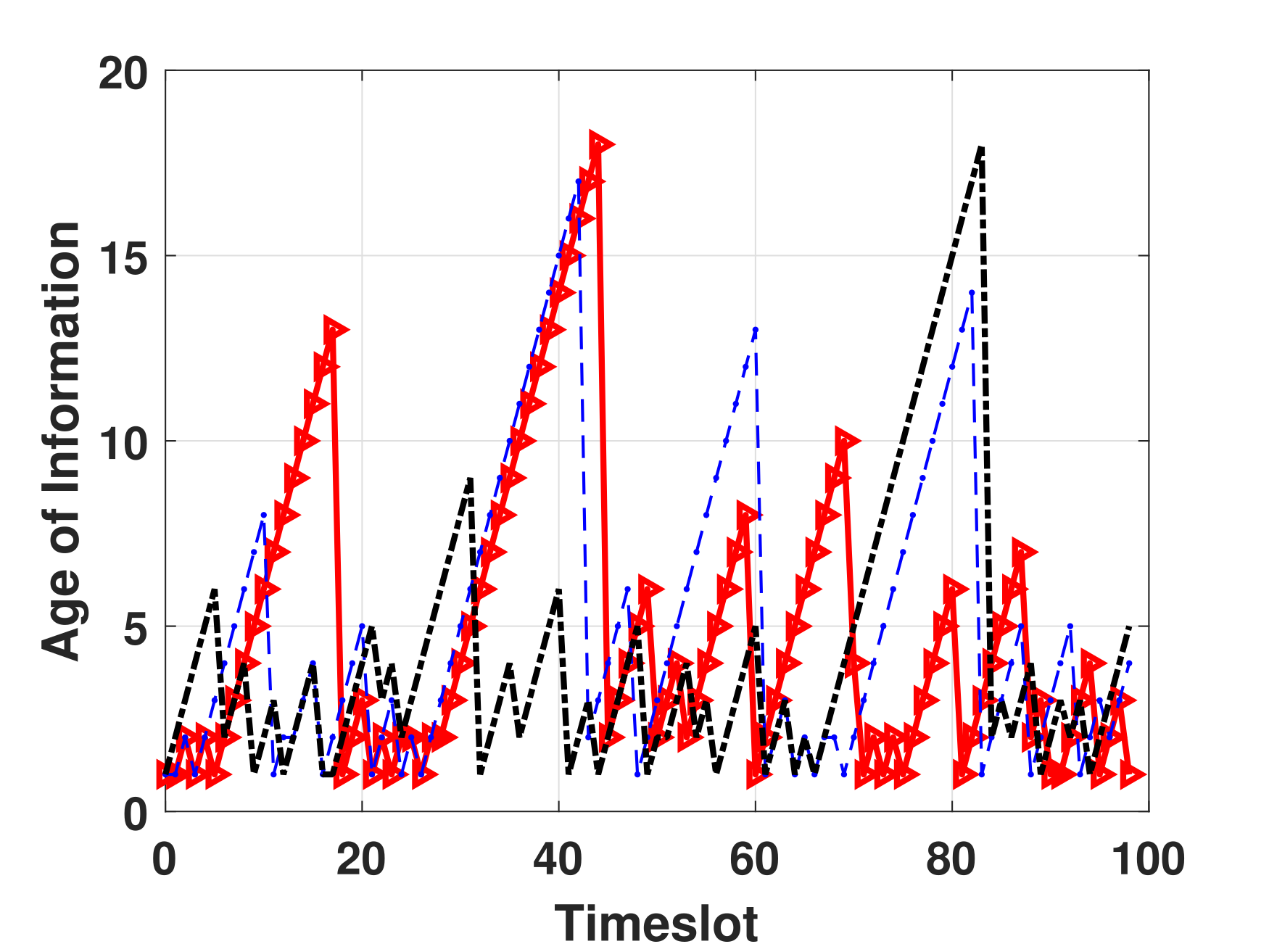}}
 \subfigure[DRL-RPM]
{\centering\includegraphics[width=0.35\textwidth]{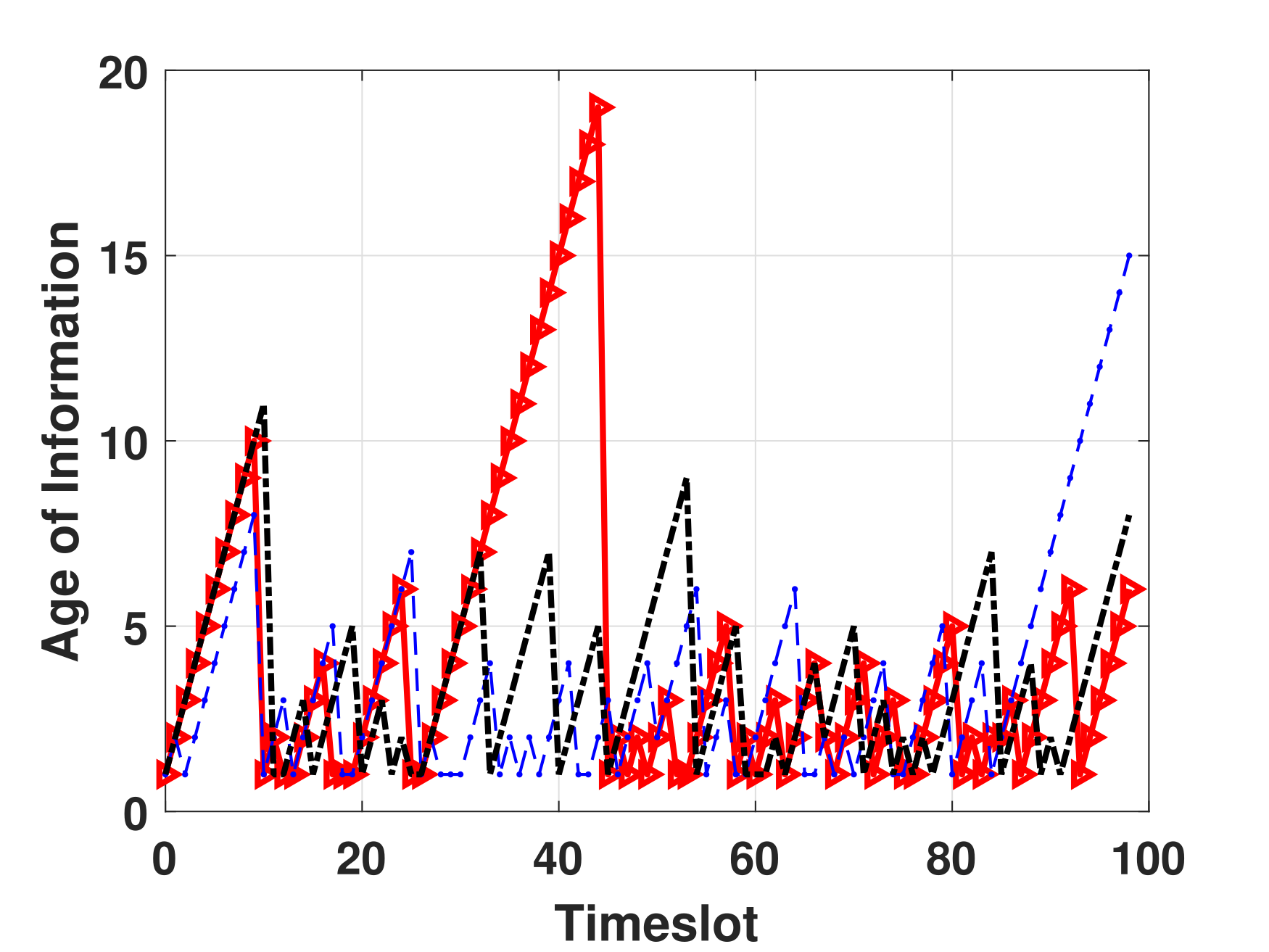}}
\caption{The performance comparison of different algorithms for a sample of three traffic streams.}
\label{largeinst}
\end{figure*}
\par {Fig. \ref{AgeVsRISdistance} illustrates the impact of the RIS location from the perspective of the BS and the destinations.  As delineated by Fig. \ref{AgeVsRISdistance}, the distance from the BS to the RIS, $d_{\rm B,R}$ is varied starting by placing it next to the BS (at $1$m distance) then increasing the distance up to $200$m with an increment of $50$m. Some interesting observations can be collected here. First, since the destinations are at least $200$m apart from the BS, the direct link from the BS to each destination is expected to undergo severe fading which will result in very high AoI values for the case where the  RIS is not used. The same has been experienced through simulations. Next, with the integration of the RIS, the quality of the transmitted signals can be greatly improved, which will eventually result in decreasing the AoI. It has been observed that, when the RIS is placed very close to the BS while the destinations were at least at $200$m away from the BS, the AoI values were considerably low. The reason being that is, since the direct link was not sufficient to successfully transmit the information to the respective destinations, the RIS played its role and with a well designed phase shift matrix, the resultant AoI was low for all methods that employed RIS. On top of this, our proposed approach performed pretty well as compared to other baseline approaches.}
\par {It can be seen that as the RIS is placed neither close to BS, nor close to destinations, the resultant AoI values start to increase for all approaches. Once, the RIS is installed close to the destinations, a significant improvement in terms of decreasing AoI against the No-RIS case can be provided. Again, the proposed approach outperforms the other baseline approaches. As explained earlier, the proposed approach takes advantage of the learning of the packets arrivals and also uses the SDR to efficiently configure the phase shift matrix of the RIS in order to maximize the SNR of the scheduled streams, which eventually results in reducing the AoI. To summarize, a well reasoned placement of the RIS can definitely lead to improving the overall system performance. The obtained results are in accordance with \cite{elhattab2021reconfigurable} which confirms that the best location for the RIS is either besides the BS or the users of interest.} 
\par {Finally, to understand the impact of different scheduling and phase shift optimization techniques on the AoI evolution over time, the AoI evolution is presented in Fig. \ref{largeinst} for all the algorithms. For a fair comparison, we have simulated a system where $I = 5$ traffic streams are competing to forward their information update packets to the destinations and $3$ traffic streams are selected to determine their AoI evolution over time. Fig. \ref{largeinst} depicts that the AoI evolution is substantially different for the different methods.  It can be observed that with the proposed approach, the AoI of all the streams is considerably smaller than those of the baseline methods. This is due to the fact that, as previously explained, the DRL agent learns how to schedule the traffic streams with packets to transmit such that the SNR on the selected channel is high enough to make the transmissions successful, which eventually reduces the AoI. However, the baseline approaches may undergo transmission failures due to inefficient scheduling, which results in packets' loss and re-transmission by the BS that increases the age. On the one hand, as delineated by Fig. \ref{largeinst}(b)-(d), it can be seen that baseline approaches significantly decrease the AoI for some streams. Furthermore, the AoI gets significantly increased to the maximum for other streams. This is because \textit{(i)} the RRS-SDR schedules the traffic streams in a round robin fashion irrespective of looking at the current AoI or the arrival time of the packets in the queue, \textit{(ii)} the DRL-RPM utilizes the RIS agent to learn to do scheduling but without a proper RIS configuration, which may not achieve the required SNR for the selected streams and results in poor performance, and \textit{(iii)} despite trying to schedule the streams with the worst AoI in each time-slot and properly configuring the RIS for the selected streams, the GS-SDR is limited due to the fact that it does scheduling attempt without having any knowledge about the arrival of packets. That's why, optimizing the RIS phase shifts alone may not guarantee that the scheduled stream would also have a packet to transmit and would increase the AoI. To conclude, our proposed approach outperforms all the baseline methods.} 
\section{Conclusion}
\label{conclusion}
{In this paper, we have investigated the role of RIS in transmitting the status update messages of multiple traffic streams to their respective destinations in order to keep the information fresh. The time-stamped status-update messages arrive to the BS with a stochastic arrival process and are selected following a scheduling policy to be forwarded to their respective destinations with the aid of RIS.  We have formulated an optimization problem to find the efficient scheduling policy that minimizes the expected sum AoI evaluating the combined impact of stochastic packet arrivals, scheduling policy and RIS phase shift. The formulated optimization problem is
a mixed integer non-convex optimization problem, which is difficult to solve. To circumvent the high-coupled optimization variables, we decompose the original problem into an outer traffic stream scheduling problem and an inner RIS phase-shift matrix problem. For the outer problem, owing to its complexity and stochastic nature of packet arrivals, we resort to deep reinforcement learning solution where the traffic stream scheduling is modeled as a MDP, and PPO is invoked to solve it. On the other hand, the inner problem to determine the RIS configuration is solved through SDR. Numerical results demonstrate the effectiveness of the proposed algorithm, which was also verified through extensive comparisons with other algorithmic solutions. 
%For future works, we can extend the current framework to consider the impact of other channel access techniques to further enhance the network capacity which is a point of interest for future wireless networks. 
}   
 
%\clearpage
 \bibliographystyle{IEEEtran}
 \bibliography{IEEEabrv,reference}

\end{document}